\documentclass[onecolumn,a4paper,accepted=2026-07-29]{quantumarticle}
\pdfoutput=1 

\usepackage{graphicx}
\usepackage{physics}
\usepackage{amsmath}
\usepackage{amssymb}
\usepackage{amsthm}
\usepackage{bbold}
\usepackage{algorithm}
\usepackage{algpseudocode}
\usepackage{array}
\usepackage{booktabs}
\usepackage{tabularx}
\usepackage{hyperref}

\usepackage{xcolor}
\newcommand{\rev}[1]{#1}

\usepackage{chngcntr}
\usepackage{apptools}
\AtAppendix{\counterwithin{lemma}{section}}

\theoremstyle{definition}
\newtheorem{theorem}{Theorem}
\newtheorem{lemma}{Lemma}

\newcommand{\dnorm}[1]{\left\|#1\right\|_{\diamond}}
\newcommand{\Lk}[1]{\hat{\mathcal{L}}_{#1}}

\title{Faster Quantum Simulation Of Markovian Open Quantum Systems Via Randomisation}

\author{I. J. David}
\affiliation{Discipline of Physics, School of Agriculture and Science, University of KwaZulu-Natal, Durban 4001, South Africa.}
\affiliation{National Institute for Theoretical and Computational Sciences (NITheCS), KZN Node, University of KwaZulu-Natal, Durban 4001, South Africa.}

\author{I. Sinayskiy}
\affiliation{Discipline of Physics, School of Agriculture and Science, University of KwaZulu-Natal, Durban 4001, South Africa.}
\affiliation{National Institute for Theoretical and Computational Sciences (NITheCS), KZN Node, University of KwaZulu-Natal, Durban 4001, South Africa.}

\author{F. Petruccione}
\affiliation{National Institute for Theoretical and Computational Sciences (NITheCS), Stellenbosch, South Africa.}
\affiliation{School of Data Science and Computational Thinking, Department of Physics, Stellenbosch University, Stellenbosch 7604, South Africa.}

\begin{document}
\maketitle

\begin{abstract}
 When simulating the dynamics of open quantum systems with quantum computers, it is essential to accurately approximate the system's behaviour while preserving the physicality of its evolution. Traditionally, for Markovian open quantum systems, this has been achieved using first and second-order Trotter-Suzuki product formulas or probabilistic algorithms. In this work, we introduce novel non-probabilistic algorithms for simulating Markovian open quantum systems using randomisation. Our methods, including first and second-order randomised Trotter-Suzuki formulas and the QDRIFT channel, not only maintain the physicality of the system's evolution but also enhance the scalability and precision of quantum simulations. We derive error bounds and step count limits for these techniques, bypassing the need for the mixing lemma typically employed in Hamiltonian simulation proofs.
Furthermore, we implement these randomised algorithms using Classical Sampling (CS), demonstrating their gate complexity advantages over deterministic TS product formulas. This work systematically extends powerful randomisation techniques from Hamiltonian simulation to the general setting of Markovian open quantum systems, highlighting their potential to enable faster and more accurate simulations.
\end{abstract}

\newpage
\tableofcontents
\newpage

\section{Introduction}

The simulation of quantum dynamics is essential for understanding large and complex quantum systems. The exponential scaling of resources required by classical methods prompted Feynman and Manin to propose using quantum computers to simulate quantum phenomena directly \cite{feynman1982simulating,manin1980computable}. This led to the development of several algorithms for simulating both closed and open quantum systems that leverage quantum mechanical principles, such as superposition, entanglement, and quantum parallelism. These algorithms provide a clear computational advantage over classical algorithms. 

The earliest algorithm for simulating the dynamics of closed quantum systems was introduced by Lloyd \cite{lloyd1996universal}. This type of simulation, known as Hamiltonian simulation, aims to construct an approximation of a unitary evolution generated by a system's Hamiltonian. This approximation is efficiently implemented on a quantum computer up to a chosen precision, with the efficiency being primarily determined by the number of quantum gates required.

The most prevalent method for Hamiltonian simulation employs Trotter-Suzuki (TS) product formulas \cite{berry2007efficient,suzuki1990fractal,suzuki1991general} to approximate the unitary evolution generated by the Hamiltonian. Extensive research has been conducted on these TS product formulas \cite{childs2018toward,childs2021theory} to evaluate their efficiency in simulating quantum dynamics. In addition to TS product formulas, other novel quantum algorithms have been developed that enhance precision and improve the gate complexity when compared to TS product formulas \cite{campbell2019random, childs2012hamiltonian, childs2019faster, childs2019nearly, hagan2023composite, kieferova2019simulating, low2017optimal, low2019hamiltonian, berry2020time, berry2015simulating} . Notable among these algorithms are Linear Combination of Unitaries (LCU) \cite{childs2012hamiltonian}, Quantum Signal Processing (QSP) \cite{low2017optimal,low2019hamiltonian}, truncated Taylor series \cite{berry2015simulating}, and randomisation-based approaches such as randomised TS product formulas \cite{childs2019faster} and QDRIFT \cite{campbell2019random}.

Significant progress has also been made in the development of algorithms for simulating open quantum systems (OQS), despite the unique challenges that arise. An OQS is characterised by its interaction with the environment, allowing for the exchange of energy and information \cite{breuer2002theory}. This work focuses on the simulation of Markovian OQS, where systems exhibit no memory effects and the dynamics are governed by the Gorini-Kossakowski-Sudarshan-Lindblad (GKSL) master equation \cite{gorini1976completely,lindblad1976generators}. The dynamical evolution of a Markovian OQS is represented by a quantum channel (or dynamical map), which is a Completely Positive and Trace Preserving (CPTP) map generated by the GKSL generator. Efficiently simulating the dynamics of a Markovian OQS on a quantum computer requires approximating the quantum channel that describes the system's evolution while preserving its CPTP nature and maintaining the physicality of the evolution. 

First and second-order TS product formulas have been utilised to simulate OQS \cite{sweke2015universal}, as they guarantee that the approximations are CPTP maps. However, higher-order TS product formulas are infeasible due to their recursive construction which results in non-CPTP maps. Alternative algorithms for simulating OQS have been proposed \cite{li2022simulating, suri2023two,childs2016efficient,cleve2016efficient}, most of which are probabilistic, introducing a non-zero probability of failure. Despite their precision and gate complexity improvements, the quest for non-probabilistic algorithms for OQS simulation remains crucial. While advanced deterministic methods that achieve nearly linear scaling in simulation time have been developed \cite{low2019hamiltonian,berry2015simulating,li2022simulating, suri2023two,childs2016efficient,cleve2016efficient,pocrnic2023quantum,ding2024simulating}, they often rely on complex subroutines with significant overhead. This motivates the development of simpler, lower-overhead methods, such as the product-formula-based approaches we explore here.

Inspired by the use of randomisation for Hamiltonian simulation \cite{childs2019faster,campbell2019random}, we present two novel non-probabilistic methods for simulating Markovian OQS that make use of randomisation. Both methods maintain the physicality of the system’s evolution by producing CPTP approximations of the ideal evolution. The first method employs randomised TS product formulas up to the second order for the simulation of Markovian OQS. The randomised product formulas improve the scaling of the gate complexity with respect to the number of terms in the generator of the quantum channel, thereby facilitating faster simulation of Markovian OQS when compared to deterministic TS product formulas. The second method is inspired by the QDRIFT approach \cite{campbell2019random} for Hamiltonian simulation. In this case, the gate complexity of the QDRIFT-inspired method is independent of the number of terms in the generator, making it particularly suitable for large systems or those with numerous interacting components, such as the 2D dissipative Jaynes-Cummings model with neighbour-neighbour interaction  \cite{pena2020dynamical}. For both the randomised product formulas method and the QDRIFT method, we derive error and number of step count bounds. These bounds do not rely on the mixing lemma by Campbell and Hastings \cite{campbell2017shorter,hastings2016turning}, which applies only to Hamiltonian simulation. Our results show in all cases that the randomised algorithms have error bounds with improved dependence on the number of terms $M$ in the GKSL generator.

This paper also introduces a novel method to implement the randomised TS product formulas and the QDRIFT channel on a quantum computer using Classical Sampling (CS), which constructs a gate set by sampling from a discrete distribution on a classical computer. We demonstrate that the gate complexities for circuits constructed using CS are efficient relative to the number of terms in the generator, and they outperform the deterministic TS product formulas outlined in Section 2.

To compare these methods, we define the gate complexity as the total number of ``simple channels'' required to implement the simulation algorithm. A simple channel is considered a fundamental building block of the evolution, such as the exponential of a summand of the generator. The table below summarizes the gate complexities derived in this work, demonstrating a clear advantage over deterministic approaches in their dependence on $M$.

\begin{table}[h!]
\caption{A summary of the gate complexities of the five deterministic and randomised TS product formula methods as well as the QDRIFT channel, implemented using CS.}
\vspace{2mm}
\centering
\begin{tabular}{|>{\raggedright\arraybackslash}p{7cm}|>{\raggedright\arraybackslash}p{7cm}|}
\hline
\textbf{Method Of Simulation} & \textbf{Gate Complexity} \\
\hline
First Order TS Deterministic & $O\left((t\Lambda)^{2}M^{3}/\epsilon\right)$\\
Second Order TS Deterministic & $O\left((t\Lambda)^{3/2}M^{5/2}/\sqrt{\epsilon}\right)$\\
First Order TS Randomised (CS)& $O\left((t\Lambda)^{3/2}M^{5/2}/\sqrt{\epsilon}\right)$\\
Second Order TS Randomised (CS) & $O\left((t\Lambda)^{3/2}M^{2}/\sqrt{\epsilon}\right)$ \\
QDRIFT (CS)& $O\left((t\Gamma\Omega)^{2}/\epsilon\right)$\\
Composite Formulas \cite{borras2025quantum} (Only apples to restricted class of Lindbladians) & $O\left(t^{3/2}/\epsilon^{1/2}\right)$ \\
\rev{Higher-Order Series Expansion \cite{li2022simulating}} & \rev{$O\left(t\,\mathrm{poly}(\log(t/\epsilon))\right)$}\\
\rev{Repeated Interactions (Iterative Qubitization) \cite{pocrnic2023quantum}} & \rev{$O\left(\alpha_{0}t + n_{\mathrm{RI}}\log(1/\epsilon)/\log\log(1/\epsilon)\right)$}\\
\rev{Repeated Interactions ($2k$-th Order Trotter-Suzuki) \cite{pocrnic2023quantum}} & \rev{$O\left(\dfrac{\left(\alpha_{0}t+\alpha_{I}\sqrt{n_{\mathrm{RI}} t}\right)^{1+1/(2k)}}{n_{\mathrm{RI}}\,\epsilon^{1/(2k)}}\right)$}\\
\hline
\end{tabular}
\label{Table1}
\end{table}

In Table \ref{Table1} we see that the randomised formulas (CS) and the QDRIFT channel all improve on the gate complexities dependence on $M$ when compared to the deterministic TS product formulas. The first order randomised TS formula (CS) scales the same as the second order deterministic formula. This is an improvement over the first order deterministic TS formula as we see an improvement in the scaling of the gate complexity with respect to the number of terms in the generator $M$, as it now scales as $M^{5/2}$ instead of $M^{3}$. The second order randomised TS (CS) gives us a quadratic dependence on $M$ which is much better than both the deterministic TS product formulas and even the first order randomised TS formula. For the QDRIFT channel (CS), we observe that there is no dependence on $M$ in the gate complexity. This makes this method ideal for systems with a large number of terms $M$. For example, two and three dimensional Jaynes-Cummings models with neighbour-neighbour interactions \cite{pena2020dynamical} and two dimensional Heisenberg models with boundary driving and local dissipation \cite{vznidarivc2014large}. The gate complexity of the QDRIFT channel (CS) depends on the total rate $\Gamma = \sum_{k=1}^{M}\gamma_{k}$ rather than explicitly on the number of generator terms $M$. This is not an unconditional improvement: since $\Gamma = \sum_{k=1}^{M}\gamma_{k} \leq M\max_{k}\gamma_{k}$, the total rate grows linearly with $M$ whenever the dissipation rates are comparable, in which case the QDRIFT channel and the deterministic TS product formulas scale comparably in $M$. The advantage of QDRIFT is therefore realised for systems in which $\Gamma$ is small or remains bounded independently of $M$ -- for example systems dominated by a few strong dissipators, or where the summed rates do not grow with the number of generator terms. \rev{In the two repeated-interaction entries of Table \ref{Table1}, $\alpha_{0}$ and $\alpha_{I}$ denote the coefficient $1$-norms of the system-plus-bath and interaction Hamiltonians respectively, and $n_{\mathrm{RI}} \in O(t^{2}\left\|\mathcal{L}\right\|_{1\to1}^{2}/\epsilon)$ is the number of interactions required to reach accuracy $\epsilon$ in the weak-coupling limit (denoted $\nu$ in \cite{pocrnic2023quantum}).} \rev{We stress that the entries of Table \ref{Table1} are not stated in a common access model and are therefore not directly comparable. The deterministic bounds quoted from \cite{li2022simulating,pocrnic2023quantum} are expressed in oracular models, in which part of the dependence on $M$ is absorbed into the cost of constructing and applying a block encoding of the generator; the absence of an explicit $M$ from such a bound therefore does not by itself imply that the compiled cost is independent of $M$. Our own complexities instead count simple channels, so that although the QDRIFT channel (CS) carries no explicit $M$-scaling, the circuit overhead of realising each simple channel is not counted by this metric. Table \ref{Table1} should accordingly be read as a summary of how each method scales within its own cost model, and not as a like-for-like ranking.}

Our approaches offer distinct advantages and trade-offs, making them suitable for different simulation scenarios, as summarized in Table \ref{tab:comparisons}.

\begin{table}[h!]
\caption{A summary of trade-offs and simulation scenarios of the methods developed in the paper}
\vspace{2mm}
\centering
\begin{tabularx}{\textwidth}{|>{\raggedright\arraybackslash}p{2.6cm}|>{\raggedright\arraybackslash}X|>{\raggedright\arraybackslash}X|>{\raggedright\arraybackslash}X|}
\hline
\textbf{Method} & \textbf{Pros} & \textbf{Cons} & \textbf{Best Use Case}\\
\hline
Randomised TS (CS) & \rev{Better scaling in the number of generator terms $M$ than deterministic TS; the first order formula matches the second order deterministic TS.} & \rev{Requires a classical co-processor for sampling.} & \rev{General-purpose improvement over the deterministic formulas, particularly at longer simulation times.}\\
\hline
QDRIFT (CS) & \rev{Gate complexity independent of $M$.} & \rev{Gate complexity quadratic in the simulation time $t$.} & \rev{Systems with many generator terms $M$, simulated for short times.}\\
\hline
\end{tabularx}
\label{tab:comparisons}
\end{table}

Concurrent with our work, other methods have recently been proposed that also leverage randomisation for the simulation of open quantum systems \cite{borras2025quantum,chen2024randomized}. For example, a qDRIFT-type approach was developed where the Lindbladian is decomposed into an ensemble of simpler generators, with analysis provided for both average and typical realizations \cite{chen2024randomized}. Another approach combines a second-order product formula with randomized compiling of the dissipative dynamics, achieving improved scaling for a restricted but physically relevant class of Lindbladians \cite{borras2025quantum}.
Our work complements these developments by providing a direct and general extension of both the randomised Trotter-Suzuki formulas and the QDRIFT protocol to arbitrary Markovian dynamics governed by the GKSL equation. A key distinction of our analysis is the derivation of rigorous error bounds that do not rely on the mixing lemma, which is specific to Hamiltonian simulation. Furthermore, we detail a concrete implementation scheme for our protocols using classical sampling.

The structure of this paper is as follows: Section 2 provides the preliminary knowledge necessary for understanding Markovian OQS and background information on deterministic TS product formulas for simulating these systems. Section 3 details the methodology for using first-order randomised TS product formulas to approximate OQS evolution and computes the associated precision and step bounds. Section 4 discusses the second-order randomised TS product formula. Section 5 covers the QDRIFT method for simulating Markovian OQS.
Section 6 describes the implementation of these methods using CS on a quantum computer and derives the resulting gate complexities, comparing them with those of the deterministic TS product formulas. Finally, Section 7 summarises the findings and presents concluding remarks.

\section{Preliminaries}
\label{section2}

In this section we shall provide the necessary background information on OQS and quantum channels. We will also recall some basic definitions and results for the quantum simulation of Markovian OQS using deterministic TS product formulas.
 
\subsection{Background}
The state space of a $d$-dimensional quantum system is $\mathcal{H}_{s} \cong \mathbb{C}^{d}$. The quantum state of such a system is described by a density operator $\rho \in \mathcal{S(H}_{s}) \subset \mathcal{B}(\mathcal{H}_{s})$, where $\mathcal{S(H}_{s})$ is the space of states on the Hilbert space. Specifically, $\mathcal{S(H}_{s})$ is the space of operators on $\mathcal{H}_{s}$ that satisfy,
\begin{align}
\label{eq1}
    \rho \geq 0,  && \mathrm{tr}(\rho)=1, && \rho=\rho^{\dagger},
\end{align}

\noindent and $\mathcal{B}(\mathcal{H}_{s})$ is the set of all bounded linear operators acting on the Hilbert space $\mathcal{H}_{s}$. The space of states $\mathcal{S(H}_{s})$ can have a matrix representation so that the density operators can be represented by $d\times d$ matrices which satisfy the properties in (\ref{eq1}). Quantum channels provide a general framework for describing the evolution of quantum states. These are Completely Positive and Trace Preserving (CPTP) maps \cite{breuer2002theory},

\begin{align}
\label{eq2}
   T:\mathcal{S(H}_{s})\rightarrow \mathcal{S(H}_{s}).
\end{align}

\noindent However we are interested only in Markovian continuous time evolution, which is described by a continuous single parameter semigroup of quantum channels $\{T_{t}\}$ which satisfy:

\begin{align}
\label{eq3}
   T_{t}T_{s}=T_{t+s}, \hspace{10mm} T_{0}=\mathbb{1} \hspace{10mm} t,s \in \mathbb{R}_{+}.
\end{align}

Also if we introduce time dependence to the state of our system then the density matrix, $\rho(t)$ which describes the quantum state at some time $t \geq 0$, can be written as

\begin{align}
\label{eq4}
   \rho(t)=T_{t}(\rho(0))=T_{t}\rho(0).
\end{align}

\noindent Every semigroup $\{T_{t}\}$ has generator $\mathcal{L}$ such that,

\begin{align}
\label{eq5}
   T_{t}=e^{t\mathcal{L}}=\sum_{j=0}^{\infty}\frac{t^{j}}{j!}\mathcal{L}^{j},
\end{align}

\noindent where $\mathcal{L}$ satisfies the master equation,
\begin{align}
\label{eq6}
   \frac{d}{dt}\rho(t)=\mathcal{L}(\rho(t)).
\end{align}

\noindent The generator $\mathcal{L}$ is the generator of a continuous one parameter Markovian semigroup $\{T_{t}\}$ if and only if it can be written in the celebrated Gorini-Kossakowski-Sudarshan-Lindblad (GKSL) form \cite{gorini1976completely,lindblad1976generators},

\begin{align}
\label{eq7}
   \mathcal{L}(\rho)= -i[H,\rho] +\sum_{k=2}^{M}\gamma_{k}\left(L_{k}\rho L_{k}^{\dagger}-\frac{1}{2}\left\{ L_{k}^{\dagger}L_{k},\rho \right\} \right),
\end{align}

\noindent where $H=H^{\dagger} \in \mathcal{M}_{d}(\mathbb{C})$ is the Hamiltonian and $\mathcal{M}_{d}(\mathbb{C})$ is the set of all $d\times d$ matrices with complex entries, $\gamma_{k}\geq 0$ are the decay rates, $L_{k}\in \mathcal{M}_{d}(\mathbb{C})$ are called the jump operators and $M=d^{2}$.
\\
It will be useful to write the generator in a more compact form,

\begin{align}
\label{eq8}
   \mathcal{L}(\rho)=\mathcal{L}_{1}(\rho)+\sum_{k=2}^{M}\gamma_{k}\mathcal{L}_{k}(\rho)=\sum_{k=1}^{M}\gamma_{k}\mathcal{L}_{k}(\rho),
\end{align}

\noindent where $\gamma_{1}=1$ and,

\begin{align}
\label{eq9}
   \mathcal{L}_{1}(\rho)=-i[H,\rho], && \mathcal{L}_{k}(\rho)=L_{k}\rho L_{k}^{\dagger}-\frac{1}{2}\left\{ L_{k}^{\dagger}L_{k},\rho \right\},
\end{align}

\noindent for $k=2,...,M$. Sometimes it will be convenient to absorb the decay rates $\gamma_{k}$ into each $\mathcal{L}_{k}$, by defining $\hat{\mathcal{L}}_{k}=\gamma_{k}\mathcal{L}_{k}$ the generator can be written as

\begin{align}
\label{eq10}
   \mathcal{L}(\rho)=\sum_{k=1}^{M}\hat{\mathcal{L}}_{k}(\rho).
\end{align}
Throughout this work we will need to construct approximations of a quantum channel and measure the precision of our approximation to the ideal quantum channels. Since quantum channels are superoperators which act on the space of operators we need to use a superoperator norm. This work will make use of the diamond norm as a measure of the precision of our approximation \cite{watrous2009semidefinite}. The diamond norm of a superoperator $V: \mathcal{B(H}_{s}) \rightarrow \mathcal{B(H}_{s})$ is
\begin{align}
   \label{eq11}
   \dnorm{V}=\sup_{A; \|A\|_{1}=1}\|(V\otimes \mathbb{1})(A)\|_{1},
\end{align}
where $\mathbb{1}: \mathcal{B(H}_{s}) \rightarrow \mathcal{B(H}_{s})$ is the identity superoperator, $A \in \mathcal{B(H}_{s})$ is an operator and $\| \cdot \|_{1}$ is the trace norm and it is defined as
\begin{align}
   \|A\|_{1}=\mathrm{tr}(\sqrt{AA^{\dagger}}),
\end{align}
for some operator $A$. In most of the literature on simulating open quantum systems \cite{sweke2015universal, childs2016efficient}, the $1\rightarrow 1$ Schatten norm \cite{watrous2004notes} is used as the measure of the precision of the approximation to the ideal quantum channel. However, the diamond norm improves over the $1\rightarrow 1$ Schatten norm as it takes into account entanglement with respect to a reference system. Using the diamond norm we can immediately find an upper-bound on the generator $\mathcal{L}$ in equation (\ref{eq10}):
\begin{align}    \dnorm{\mathcal{L}}&=\dnorm{\sum_{k=1}^{M}\Lk{k}}\leq \sum_{k=1}^{M}\dnorm{\Lk{k}}.
\end{align}
If we define $\Lambda:=\max_{k}\left\{\dnorm{\Lk{k}}\right\}$ then,
\begin{align}
   \label{boundongenerator}
   \dnorm{\mathcal{L}}\leq \sum_{k=1}^{M}\dnorm{\Lk{k}}\leq \sum_{k=1}^{M}\Lambda =M\Lambda.
\end{align}
We emphasise that the objects $\hat{\mathcal{L}}_{k}$ (and the rate-normalised $\mathcal{L}_{k}$) appearing in $\Lambda$ are the \emph{generators} of the dynamics --- that is, the GKSL summands defined in equations (\ref{eq9})--(\ref{eq10}). These are Hermiticity-preserving and trace-annihilating superoperators, but they are \emph{not} quantum channels: they are not trace-preserving as maps, and consequently their diamond norms are in general not equal to one. The completely positive and trace-preserving (CPTP) maps that actually appear in our circuits are the exponentials $e^{\tau\hat{\mathcal{L}}_{k}}$, which do have unit diamond norm. The parameter $\Lambda := \max_{k}\dnorm{\hat{\mathcal{L}}_{k}}$ is therefore a finite, non-trivial quantity that scales with the magnitude of the Hamiltonian and the jump operators (and, since $\hat{\mathcal{L}}_{k}=\gamma_{k}\mathcal{L}_{k}$, with the decay rate $\gamma_{k}$).
The diamond norm can be related to the trace norm via the following inequality. Given two superoperators $V$ and an operator $A$ we have by definition:
\begin{align}
   \label{statenormineq}
   \norm{A}_{1}\leq \dnorm{V}.
\end{align}
This inequality will play an important role in bounding the distances between states in quantum simulation. The following lemma, the proof of which can be found in Appendix \ref{AppendixA}, will help derive error bounds between a quantum channel $T_{t}$ and its approximation.
\begin{lemma}
\label{lemma1}
   Given two quantum channels $T$ and $V$ and some positive integer $N$,
   \begin{align}
       \label{eq12}
       \dnorm{T^{N}-V^{N}}\leq N\dnorm{T-V}.
   \end{align}
\end{lemma}

\noindent Now that we have outlined some background information about OQS, in the next sub-section we will briefly describe digital quantum simulation and show how we can simulate Markovian OQS using deterministic TS product formulas.

\subsection{Deterministic Digital Simulation of Markovian Open Quantum Systems}

\noindent The main goal of digital quantum simulation of Markovian OQS is to find novel ways of constructing an approximation $\tilde{T}_{t}$ of the total evolution $T_{t}=\exp(t\mathcal{L})$ such that for a given GKSL generator $\mathcal{L}$, a precision $\epsilon >0$, a simulation time $t\geq 0$ and a distance measure dist$(\cdot,\cdot)$,
\begin{align}
	\mathrm{dist}(T_{t},\tilde{T}_{t})\leq \epsilon
\end{align}
where $\tilde{T}_{t}$ could be implemented on a quantum computer efficiently. The most common way to obtain this approximation is by using Trotter-Suzuki (TS) product formulas \cite{suzuki1990fractal,suzuki1991general}.
Using TS product formulas, the total evolution $T_{t}=\exp(t\sum_{k=1}^{M}\hat{\mathcal{L}}_{k})$ is approximated as some product of simpler channels up to a precision $\epsilon\geq 0$ when using the diamond norm as the distance measure for the quantum channels. For the purpose of this work, we define a simple channel as a fundamental building block of our simulation circuits. These include the constituent channels of the form $\exp(\tau \Lk{k})$ derived from the GKSL generator, as well as auxiliary operations like controlled-SWAP channels required for certain randomised implementations. The specific circuit implementation of these simple channels is generator-dependent and considered out of scope for our complexity analysis. However, for many physical systems where the generator terms $\Lk{k}$ are composed of local operators (e.g., Pauli strings), efficient quantum circuits for these channels can be constructed using standard unitary dilation techniques, such as the Stinespring representation \cite{stinespring1955positive,wolf2012quantum}.

The same approach was introduced in \cite{sweke2015universal},  however, the $1\rightarrow 1$ Schatten norm was used instead of the diamond norm. We start by dividing the time $t \geq 0$ into $N\in \mathbb{N}$ steps so that we have a small time step $\tau =t/N$. Next we construct approximations of the total channel $T_{t}$ using TS product formulas that approximate the channel for a small time step $T_{\tau}$ and then simulate the TS product formula $N$ times. For example, we can approximate the evolution $T_{\tau}$ up to first order by using the following deterministic TS product formula,
\begin{align}
   S_{1}^{(det)}(\tau)=\prod_{k=1}^{M}e^{\tau\hat{\mathcal{L}}_{k}},
\end{align}
where we shall refer to the exponentials of the form $\exp(\tau\Lk{k})=\exp(\tau\gamma_{k}\mathcal{L}_{k})$ as constituent channels. Similarly we can approximate $T_{\tau}$ up to second order using the formula,
\begin{align}
   S_{2}^{(det)}(\tau)=\prod_{k=1}^{M}e^{\frac{\tau}{2}\hat{\mathcal{L}}_{k}}\prod_{k'=M}^{1}e^{\frac{\tau}{2}\hat{\mathcal{L}}_{k'}}.
\end{align}
We refer to these product formulas as deterministic because the ordering of the exponentials in $S_{1}^{(det)}$ and $S_{2}^{(det)}$ is known before hand. We state two theorems that outline how  TS product formulas are used to approximate the total evolution $T_{t}$.

\begin{theorem}{(First Order Deterministic TS Product Formula):}
\label{theorem1}
   Given the generator $\mathcal{L}$, as in equation (\ref{eq10}), of a quantum channel $T_{t}$ and some time $t\geq 0$. Define the first order deterministic TS product formula as,
   \begin{align}
       \label{eq13}
       S_{1}^{(det)}(\tau)=\prod_{k=1}^{M}e^{\tau\hat{\mathcal{L}}_{k}}.
   \end{align}
   Let $\Lambda :=\max_{k}\dnorm{\hat{\mathcal{L}}_{k}}$ and then for $N$ chosen such that $Mt\Lambda/N\leq 1$. Then,
   \begin{align}
       \label{eq14}
       \dnorm{T_{t}-S_{1}^{(det)}\left(\frac{t}{N}\right)^{N}}\leq \frac{et^{2}\Lambda^{2}M^{2}}{N},
   \end{align}
   where we choose
   \begin{align}
       \label{eq15}
       \epsilon \geq \frac{et^{2}\Lambda^{2}M^{2}}{N}, && N \geq \frac{et^{2}\Lambda^{2}M^{2}}{\epsilon}.
   \end{align}
\end{theorem}

\noindent The proof of Theorem \ref{theorem1}. can be found in Appendix \ref{AppendixA}

\begin{theorem}{(Second Order Deterministic TS Product Formula):}
\label{theorem2}
   Given the generator $\mathcal{L}$, as in equation (\ref{eq10}), of a quantum channel $T_{t}$ and some time $t\geq 0$. Define the second order deterministic TS product formula as,
   \begin{align}
       \label{eq16}
       S_{2}^{(det)}(\tau)=\prod_{k=1}^{M}e^{\frac{\tau}{2}\hat{\mathcal{L}}_{k}}\prod_{k'=M}^{1}e^{\frac{\tau}{2}\hat{\mathcal{L}}_{k'}}.
   \end{align}
   Let $\Lambda :=\max_{k}\dnorm{\hat{\mathcal{L}}_{k}}$ and we choose $N$ such that $Mt\Lambda/N \leq 1$. Then,
   \begin{align}
       \label{eq17}
       \dnorm{T_{t}-S_{2}^{(det)}\left(\frac{t}{N}\right)^{N}}\leq \frac{eM^{3}t^{3}\Lambda^{3}}{3N^{2}},
   \end{align}
where we choose
   \begin{align}
       \label{eq18}
       \epsilon \geq \frac{eM^{3}t^{3}\Lambda^{3}}{3N^{2}}, && N\geq \frac{e^{1/2}M^{3/2}t^{3/2}\Lambda^{3/2}}{\sqrt{3\epsilon}}.
   \end{align}
\end{theorem}

\begin{figure*}[h]
   \centering
   \includegraphics[scale=0.4]{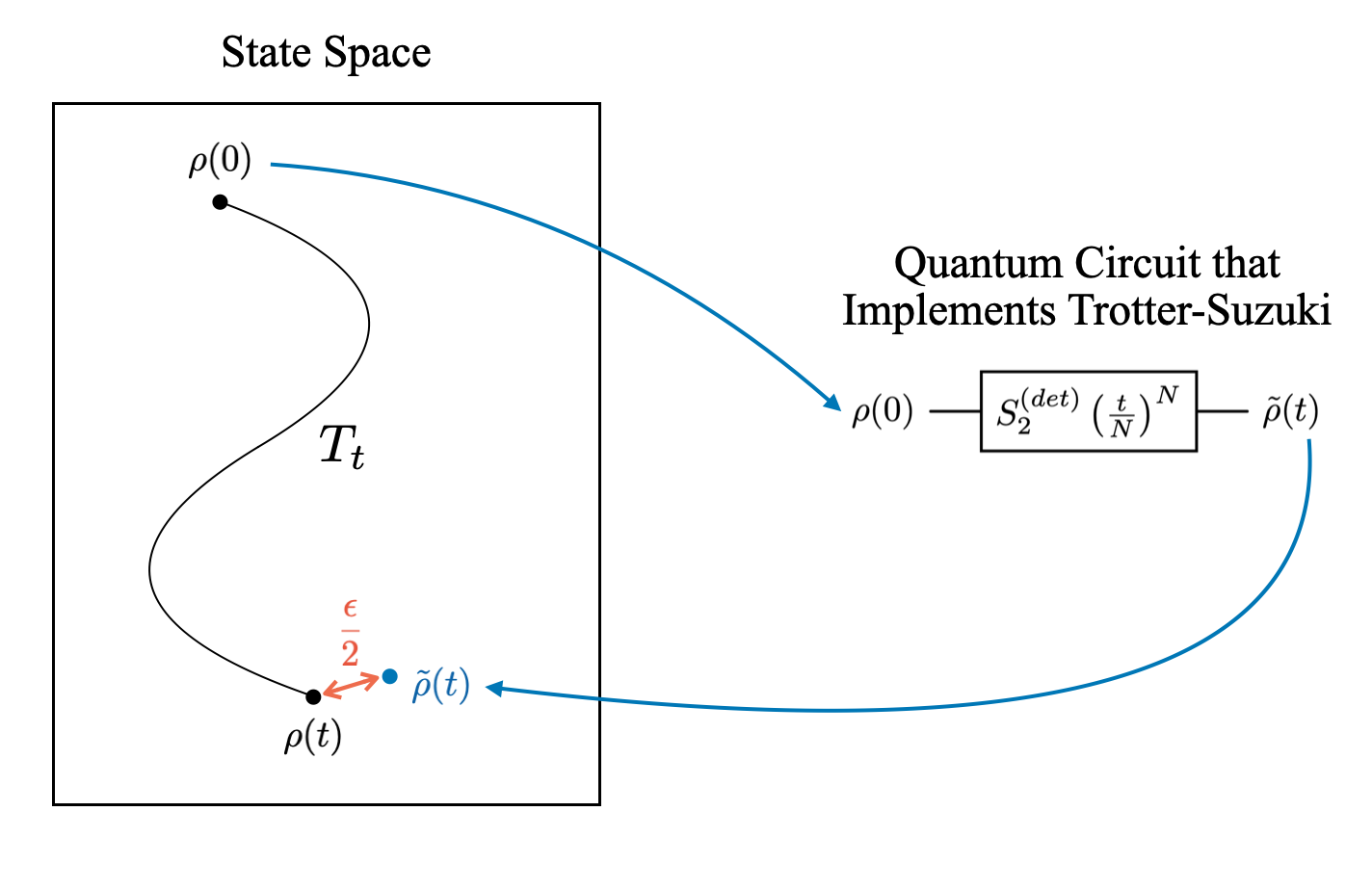}
   \caption{The key ideas behind the digital quantum simulation of an OQS are illustrated. The state $\rho(0)$ is the initial state of the system, $T_{t}$ is the quantum channel describing the systems evolution and $\rho(t)$ is the state after evolving for some time $t$. The quantum circuit implements the second order deterministic TS product formula $S_{2}^{(det)}$ which produces an output state $\tilde{\rho}(t)$ which approximates the state $\rho(t)$ up to a precision $\epsilon/2$.}
   \label{Fig1}
\end{figure*}

\noindent The proof of Theorem \ref{theorem2}. can be found in Appendix \ref{AppendixA}. Now that we have shown that we can approximate $T_{t}$ by TS product formulas, all that is left is to construct a quantum circuit that implements this product formula on a quantum computer. Since our product formula is a quantum channel, one needs to use a unitary dilation, for example the Stinespring representation of the channel \cite{stinespring1955positive,wolf2012quantum}, to construct a quantum circuit. However, in this work, when we draw quantum circuits we will only show the action of the channel on the state as this keeps the diagrams concise. To make clear how the diagrams should be interpreted, we note that wires in our circuits correspond to density matrices of a subsystem and gates correspond to quantum channels, thereafter the usual rules of quantum circuits may be inferred. As an illustrative example consider the deterministic product formula $S_{2}^{(det)}(t/N)^{N}$ that approximates the total evolution $T_{t}$ up to a precision $\epsilon$. This means that the output of the quantum circuit that implements $S_{2}^{(det)}$ is a density matrix $\tilde{\rho}(t)$ that is a distance $\epsilon/2$ from the density matrix $\rho(t)$. One can easily see this by using the definition of the trace distance between states i.e. $d_{tr}(\rho(t),\tilde{\rho}(t))$ and inequality (\ref{statenormineq}),
\begin{align}
\label{eq2-21}
   d_{tr}(\rho(t),\tilde{\rho}(t))&=\frac{1}{2}\norm{\rho(t)-\tilde{\rho}(t)}_{1}\nonumber\\
   &= \frac{1}{2}\norm{T_{t}(\rho(0))-S_{2}^{(det)}(t/N)^{N}\rho(0)}_{1}\nonumber\\
   &\leq \frac{1}{2}\dnorm{T_{t}-S_{2}^{(det)}(t/N)^N}\nonumber\\
   &\leq \frac{\epsilon}{2},
\end{align}
where $\epsilon \geq (Mt\Lambda)^{3}/3N^{2}$ as in Theorem \ref{theorem2}. Figure \ref{Fig1}. shows how we can use $S_{2}^{(det)}(t/N)^{N}$ to simulate the evolution $T_{t}$.

\noindent At this point we need to analyse the gate complexity of the quantum circuits we have constructed to implement the deterministic TS product formulas. To do this, we start by defining the gate complexity of our circuits as the number of simple channels that are implemented in each quantum circuit. For the case of deterministic TS product formulas, these simple channels are just the exponentials of the form $\exp(\tau \Lk{k})$. For example, consider the product formula $S_{1}^{(det)}(\tau)$ in (\ref{eq17}). This formula is the product of $M$ exponentials. If we consider the quantum circuit that implements $S_{1}^{(det)}(\tau)^{N}$ to approximate $\rho(t)$ to a precision
$\epsilon$, then we have $N=\left\lceil e\,t^{2}\Lambda^{2}M^{2}/\epsilon \right\rceil$ applications of $S_{1}^{(det)}(\tau)$ which implies that we have to implement $\left\lceil e\,t^{2}\Lambda^{2}M^{2}/\epsilon \right\rceil M$ exponentials, where $\lceil \cdot \rceil$ denotes the ceiling function. If we denote the gate complexity for the circuit that implements $S_{1}^{(det)}$ as $g_{1}^{(det)}$ then the complexity is given by
\begin{align}
   \label{eq2-19}
   g_{1}^{(det)}=O\left(\frac{t^{2}\Lambda^{2}M^{3}}{\epsilon}\right).
\end{align}
The second order formula $S_{2}^{(det)}(\tau)$ contains $2M$ exponentials. The quantum circuit that implements $S_{2}^{(det)}(\tau)^{N}$ to approximate $\rho(t)$ to a precision $\epsilon$ will contain $N=\left\lceil e^{1/2}\,M^{3/2}t^{3/2}\Lambda^{3/2}/\sqrt{3\epsilon} \ \right\rceil$ applications of $S_{2}^{(det)}(\tau)$. This tells us that we will have to implement $2 \left\lceil e^{1/2}\,M^{3/2}t^{3/2}\Lambda^{3/2}/\sqrt{3\epsilon}\ \right\rceil M$ exponentials. By denoting the gate complexity of this formula by $g_{2}^{(det)}$, we see that
\begin{align}
   \label{eq2-20}
   g_{2}^{(det)}=O\left(\frac{M^{5/2}t^{3/2}\Lambda^{3/2}}{\sqrt{3\epsilon}}\right).
\end{align}

\section{First Order Randomised Trotter-Suzuki Formula}
\label{section3}

To define the randomised first order formula, we first need to define two useful formulas. We observe that the product in equation (\ref{eq13}), has the constituent channels $e^{\tau\Lk{k}}$ arranged from left to right starting from $e^{\tau\Lk{1}}$ and ending with $e^{\tau\Lk{M}}$. We shall call this the forward direction and define $S_{1}^{\rightarrow}(\tau)=S_{1}^{(det)}(\tau)$. We also observe that if we choose to reverse this ordering so that the product of constituent channels is arranged from right to left starting with $e^{\tau\Lk{M}}$ and ending with $e^{\tau\Lk{1}}$. We call this the reversed first order product formula and it is defined as,
\begin{align}
   \label{eq3-1}
   S_{1}^{\leftarrow}(\tau)=\prod_{k=M}^{1}e^{\tau\Lk{k}}.
\end{align}

\noindent The randomised first order Trotter-Suzuki formula can then be defined as a convex combination of first order Trotter-Suzuki formulas in both the forward and reversed orders i.e.
\begin{align}
   \label{eq3-2}
   S_{1}^{(ran)}(\tau)=\frac{1}{2}\left( S_{1}^{\rightarrow}(\tau)+ S_{1}^{\leftarrow}(\tau) \right).
\end{align}
Now consider the form of the first order deterministic Trotter-Suzuki formula i.e. $S_{1}^{(det)}$ in equation (\ref{eq13}). We know that this approximates the total channel $T_{t}$ up to first order with a second order error term. However, using the formula $S_{1}^{(ran)}$, we shall see that we obtain an improvement in both the precision $\epsilon$ and the number of steps $N$.  The following theorem shows the error bound and gate complexity for the first order randomised Trotter-Suzuki formula. It should be noted that this result has been proven for the simulation of closed quantum systems (Hamiltonian simulation) \cite{childs2019faster}, however the proof relies on the mixing lemma developed by Campbell and Hastings \cite{campbell2017shorter,hastings2016turning}. Since this lemma is not applicable to open quantum systems, we present a proof for the error bound and complexity of the first order randomised formula that does not rely on the mixing lemma.

\begin{theorem}
\label{theorem3}
   Given the generator $\mathcal{L}$, as in equation (\ref{eq10}), of a quantum channel $T_{t}$ and some time $t\geq 0$.  Define the first order randomised product formula as in equation (\ref{eq3-2}). Let  $\Lambda :=\max_{k}\dnorm{\hat{\mathcal{L}}_{k}}$ and choose $N$ such that $Mt\Lambda/N\leq 1$. Then,
   \begin{align}
       \label{eq3-3}
       \dnorm{T_{t}-S_{1}^{(ran)}\left(\frac{t}{N}\right)^{N}}\leq \frac{eM^{3}t^{3}\Lambda^{3}}{3N^{2}},
   \end{align}
   where
   \begin{align}
       \label{eq3-4}
       \epsilon \geq \frac{eM^{3}t^{3}\Lambda^{3}}{3N^{2}},
   \end{align}
\begin{align}
   \label{eq3-5}
   N\geq \frac{e^{1/2}(t\Lambda M)^{3/2}}{\sqrt{3\epsilon}}.
\end{align}
\end{theorem}

\begin{proof}
   Making use of Lemma 1. we can write,
   \begin{align}
       \label{eq3-7}
       \dnorm{T_{t}-S_{1}^{(ran)}\left(\frac{t}{N}\right)^{N}}&=\dnorm{T_{\frac{t}{N}}^{N}-S_{1}^{(ran)}\left(\frac{t}{N}\right)^{N}}\\
       \label{eq3-7a}
       &\leq N\dnorm{T_{\tau}-S_{1}^{(ran)}\left(\tau\right)},
   \end{align}
where $\tau=t/N$. Now we can bound $\dnorm{T_{\tau}-S_{1}^{(ran)}\left(\tau\right)}$, we start by performing a Taylor expansion of $T_\tau$ and writing out explicitly the terms up to second order,
\begin{align}
   \label{eq3-8}
   T_{\tau}=\exp(\tau\sum_{k=1}^{M}\Lk{k})&=\mathbb{1}+\tau\sum_{j=1}^{M}\Lk{j}+\frac{\tau^{2}}{2}\left(\sum_{j=1}^{M}\Lk{j}\right)^{2}+\sum_{n=3}^{\infty}\frac{\tau^{n}}{n!}\mathcal{L}^{n},\\
   &=\mathbb{1}+\tau\sum_{j=1}^{M}\Lk{j}+\frac{\tau^{2}}{2}\sum_{j=1}^{M}\Lk{j}^{2}+\frac{\tau^{2}}{2}\sum_{\substack{j,k=1\\j\neq k}}^{M}\Lk{j}\Lk{k}+\sum_{n=3}^{\infty}\frac{\tau^{n}}{n!}\mathcal{L}^{n}.
\end{align}
Then, if we consider $S_{1}^{(ran)}(\tau)=\frac{1}{2}\left( S_{1}^{\rightarrow}(\tau)+ S_{1}^{\leftarrow}(\tau) \right)$, we Taylor expand $S_{1}^{\rightarrow}(\tau)$ and $S_{1}^{\leftarrow}(\tau)$,
\begin{align}
   \label{eq3-9}
   S_{1}^{\rightarrow}(\tau)&=\prod_{k=1}^{M}e^{\tau\Lk{k}},\\
   &=\sum_{j_{1},...,j_{M}=0}^{\infty}\frac{\tau^{j_{1}+...+j_{M}}}{j_{1}!...j_{M}!}\Lk{1}^{j_{1}}\Lk{2}^{j_{2}}...\Lk{M}^{j_{M}},\\
   &=\sum_{p=0}^{\infty}\sum_{\substack{j_{1},...,j_{M}=0\\ \sum_{\mu}j_{\mu}=p}}^{p}\frac{\tau^{j_{1}+...+j_{M}}}{j_{1}!...j_{M}!}\Lk{1}^{j_{1}}\Lk{2}^{j_{2}}...\Lk{M}^{j_{M}},\\
   &=\mathbb{1}+\tau\sum_{j=1}^{M}\Lk{j}+\frac{\tau^{2}}{2}\sum_{j=1}^{M}\Lk{j}^{2}+\tau^{2}\sum_{\substack{k,l=1\\ k<l}}^{M}\Lk{k}\Lk{l}+\sum_{p=3}^{\infty}\sum_{\substack{j_{1},...,j_{M}=0\\ \sum_{\mu}j_{\mu}=p}}^{p}\frac{\tau^{j_{1}+...+j_{M}}}{j_{1}!...j_{M}!}\Lk{1}^{j_{1}}\Lk{2}^{j_{2}}...\Lk{M}^{j_{M}},
\end{align}
and,
\begin{align}
   \label{eq3-10}
   S_{1}^{\leftarrow}(\tau)&=\prod_{k=M}^{1}e^{\tau\Lk{k}},\\
   &=\sum_{j_{1},...,j_{M}=0}^{\infty}\frac{\tau^{j_{1}+...+j_{M}}}{j_{1}!...j_{M}!}\Lk{M}^{j_{1}}\Lk{M-1}^{j_{2}}...\Lk{1}^{j_{M}},\\
   &=\sum_{p=0}^{\infty}\sum_{\substack{j_{1},...,j_{M}=0\\ \sum_{\mu}j_{\mu}=p}}^{p}\frac{\tau^{j_{1}+...+j_{M}}}{j_{1}!...j_{M}!}\Lk{M}^{j_{1}}\Lk{M-1}^{j_{2}}...\Lk{1}^{j_{M}},\\
   &=\mathbb{1}+\tau\sum_{j=1}^{M}\Lk{j}+\frac{\tau^{2}}{2}\sum_{j=1}^{M}\Lk{j}^{2}+\tau^{2}\sum_{\substack{k,l=1\\ k>l}}^{M}\Lk{k}\Lk{l}+\sum_{p=3}^{\infty}\sum_{\substack{j_{1},...,j_{M}=0\\ \sum_{\mu}j_{\mu}=p}}^{p}\frac{\tau^{j_{1}+...+j_{M}}}{j_{1}!...j_{M}!}\Lk{M}^{j_{1}}\Lk{M-1}^{j_{2}}...\Lk{1}^{j_{M}}.
\end{align}
Then $S_{1}^{(ran)}(\tau)$ can be written as
\begin{align}
   \label{eq3-11}
   S_{1}^{(ran)}(\tau)&=\mathbb{1}+\tau \sum_{j=1}^{M}\Lk{j}+\frac{\tau}{2}\sum_{j=1}^{M}\Lk{j}^{2}+\frac{\tau^{2}}{2}\sum_{\substack{k,l=1\\k<l}}^{M}\Lk{k}\Lk{l}+\frac{\tau^{2}}{2}\sum_{\substack{k,l=1\\k>l}}^{M}\Lk{k}\Lk{l}+\nonumber\\
&\frac{1}{2}\sum_{p=3}^{\infty}\sum_{\substack{j_{1},...,j_{M}=0\\ \sum_{\mu}j_{\mu}=p}}^{p}\frac{\tau^{j_{1}+...+j_{M}}}{j_{1}!...j_{M}!}\Lk{1}^{j_{1}}\Lk{2}^{j_{2}}...\Lk{M}^{j_{M}}+\frac{1}{2}\sum_{p=3}^{\infty}\sum_{\substack{j_{1},...,j_{M}=0\\ \sum_{\mu}j_{\mu}=p}}^{p}\frac{\tau^{j_{1}+...+j_{M}}}{j_{1}!...j_{M}!}\Lk{M}^{j_{1}}\Lk{M-1}^{j_{2}}...\Lk{1}^{j_{M}},\\
&=\mathbb{1}+\tau \sum_{j=1}^{M}\Lk{j}+\frac{\tau}{2}\sum_{j=1}^{M}\Lk{j}^{2}+\frac{\tau^{2}}{2}\sum_{\substack{k,l=1\\k\neq l}}^{M}\Lk{k}\Lk{l}+\nonumber\\
&\frac{1}{2}\sum_{p=3}^{\infty}\sum_{\substack{j_{1},...,j_{M}=0\\ \sum_{\mu}j_{\mu}=p}}^{p}\frac{\tau^{j_{1}+...+j_{M}}}{j_{1}!...j_{M}!}\Lk{1}^{j_{1}}\Lk{2}^{j_{2}}...\Lk{M}^{j_{M}}+\frac{1}{2}\sum_{p=3}^{\infty}\sum_{\substack{j_{1},...,j_{M}=0\\ \sum_{\mu}j_{\mu}=p}}^{p}\frac{\tau^{j_{1}+...+j_{M}}}{j_{1}!...j_{M}!}\Lk{M}^{j_{1}}\Lk{M-1}^{j_{2}}...\Lk{1}^{j_{M}}.
\end{align}
The difference between the total channel $T_{\tau}$ and $S_{1}^{(ran)}(\tau)$ is
\begin{align}
   \label{eq3-12}
   T_{\tau}&-S_{1}^{(ran)}(\tau)=\sum_{n=3}^{\infty}\frac{\tau^{n}}{n!}\mathcal{L}^{n}\nonumber\\
   &-\frac{1}{2}\sum_{p=3}^{\infty}\sum_{\substack{j_{1},...,j_{M}=0\\ \sum_{\mu}j_{\mu}=p}}^{p}\frac{\tau^{j_{1}+...+j_{M}}}{j_{1}!...j_{M}!}\Lk{1}^{j_{1}}\Lk{2}^{j_{2}}...\Lk{M}^{j_{M}}-\frac{1}{2}\sum_{p=3}^{\infty}\sum_{\substack{j_{1},...,j_{M}=0\\ \sum_{\mu}j_{\mu}=p}}^{p}\frac{\tau^{j_{1}+...+j_{M}}}{j_{1}!...j_{M}!}\Lk{M}^{j_{1}}\Lk{M-1}^{j_{2}}...\Lk{1}^{j_{M}}
\end{align}
Now, we bound the difference between $T_{\tau}$ and $S_{1}^{(ran)}(\tau)$:
\begin{align}
   \label{eq3-13}
   &\dnorm{T_{\tau}-S_{1}^{(ran)}(\tau) }\leq \sum_{n=3}^{\infty}\frac{\tau^{n}}{n!}\dnorm{\mathcal{L}^{n}}\nonumber\\
   &+\frac{1}{2}\sum_{p=3}^{\infty}\sum_{\substack{j_{1},...,j_{M}=0\\ \sum_{\mu}j_{\mu}=p}}^{p}\frac{\tau^{j_{1}+...+j_{M}}}{j_{1}!...j_{M}!}\dnorm{\Lk{1}^{j_{1}}\Lk{2}^{j_{2}}...\Lk{M}^{j_{M}}}+\frac{1}{2}\sum_{p=3}^{\infty}\sum_{\substack{j_{1},...,j_{M}=0\\ \sum_{\mu}j_{\mu}=p}}^{p}\frac{\tau^{j_{1}+...+j_{M}}}{j_{1}!...j_{M}!}\dnorm{\Lk{M}^{j_{1}}\Lk{M-1}^{j_{2}}...\Lk{1}^{j_{M}}}
\end{align}
Now, to complete the bound we need to bound the three terms in equation (\ref{eq3-13}). We start with the first term. Noting that
\begin{align}
   \label{eq3-14}
   \dnorm{\mathcal{L}^{n}}\leq \dnorm{\mathcal{L}}^{n} \leq M^{n}\Lambda^{n},
\end{align}
then
\begin{align}
   \label{eq3-15}
   \sum_{n=3}^{\infty}\frac{\tau^{n}}{n!}\dnorm{\mathcal{L}^{n}} \leq  \sum_{n=3}^{\infty}\frac{\tau^{n}}{n!}M^{n}\Lambda^{n}.
\end{align}
For the second term in equation (\ref{eq3-13}), we use the fact that the diamond norm is sub-multiplicative and $\dnorm{\Lk{i}}\leq \Lambda$ for all $i=1,...,M$ to show that
\begin{align}
   \label{eq3-16}
   \dnorm{\Lk{1}^{j_{1}}\Lk{2}^{j_{2}}...\Lk{M}^{j_{M}}}\leq \dnorm{\Lk{1}}^{j_{1}}...\dnorm{\Lk{M}}^{j_{M}}\leq \Lambda^{j_{1}+...+j_{M}}.
\end{align}
Using equation (\ref{eq3-16}) we bound the second term in equation (\ref{eq3-13}) as
\begin{align}
   \label{eq3-17}
   \frac{1}{2}\sum_{p=3}^{\infty}\sum_{\substack{j_{1},...,j_{M}=0\\ \sum_{\mu}j_{\mu}=p}}^{p}\frac{\tau^{j_{1}+...+j_{M}}}{j_{1}!...j_{M}!}\dnorm{\Lk{1}^{j_{1}}\Lk{2}^{j_{2}}...\Lk{M}^{j_{M}}}\leq \frac{1}{2}\sum_{p=3}^{\infty}\sum_{\substack{j_{1},...,j_{M}=0\\ \sum_{\mu}j_{\mu}=p}}^{p}\frac{\tau^{j_{1}+...+j_{M}}}{j_{1}!...j_{M}!}\Lambda^{j_{1}+...+j_{M}}.
\end{align}
In a similar way we find the bound for the third term in equation (\ref{eq3-13}) to be
\begin{align}
   \label{eq3-18}
   \frac{1}{2}\sum_{p=3}^{\infty}\sum_{\substack{j_{1},...,j_{M}=0\\ \sum_{\mu}j_{\mu}=p}}^{p}\frac{\tau^{j_{1}+...+j_{M}}}{j_{1}!...j_{M}!}\dnorm{\Lk{M}^{j_{1}}\Lk{M-1}^{j_{2}}...\Lk{1}^{j_{M}}}\leq  \frac{1}{2}\sum_{p=3}^{\infty}\sum_{\substack{j_{1},...,j_{M}=0\\ \sum_{\mu}j_{\mu}=p}}^{p}\frac{\tau^{j_{1}+...+j_{M}}}{j_{1}!...j_{M}!}\Lambda^{j_{1}+...+j_{M}}.
\end{align}
Now using Lemma \ref{lemmaRestrictedSum}. from Appendix \ref{AppendixB}, we can compute the restricted sums in equations (\ref{eq3-17}) and (\ref{eq3-18}). We then get the final bounds on these terms as
\begin{align}
   \label{eq3-19}
   \frac{1}{2}\sum_{p=3}^{\infty}\sum_{\substack{j_{1},...,j_{M}=0\\ \sum_{\mu}j_{\mu}=p}}^{p}\frac{\tau^{j_{1}+...+j_{M}}}{j_{1}!...j_{M}!}\dnorm{\Lk{1}^{j_{1}}\Lk{2}^{j_{2}}...\Lk{M}^{j_{M}}}\leq \frac{1}{2}\sum_{p=3}^{\infty}\frac{M^{p}\tau^{p} \Lambda^{p}}{p!},
\end{align}
and
\begin{align}
   \label{eq3-20}
    \frac{1}{2}\sum_{p=3}^{\infty}\sum_{\substack{j_{0},...,j_{M}=0\\ \sum_{\mu}j_{\mu}=p}}^{p}\frac{\tau^{j_{0}+...+j_{M}}}{j_{0}!...j_{M}!}\dnorm{\Lk{M}^{j_{1}}\Lk{M-1}^{j_{2}}...\Lk{1}^{j_{M}}}\leq  \frac{1}{2}\sum_{p=3}^{\infty}\frac{M^{p}\tau^{p}\Lambda^{p}}{p!}.
\end{align}
Substituting the bounds obtained in equations (\ref{eq3-15}), (\ref{eq3-19}) and (\ref{eq3-20}) into (\ref{eq3-13}) yields
\begin{align}
   \label{eq3-21}
   \dnorm{T_{\tau}-S_{1}^{(ran)}(\tau) }&\leq \sum_{n=3}^{\infty}\frac{M^{n}\tau^{n}\Lambda^{n}}{n!} + \sum_{p=3}^{\infty}\frac{M^{p}\tau^{p}\Lambda^{p}}{p!},\nonumber\\
   &= 2\sum_{p=3}^{\infty}\frac{M^{p}\tau^{p}\Lambda^{p}}{p!}.
\end{align}
Using Lemma F.2 from the supplementary information of \cite{childs2018toward}, which states that, for some $y \geq 0 \in \mathbb{R}$ and $k\in \mathbb{N}$
\begin{align}
   \label{eq3-22}
   \sum_{n=k}^{\infty}\frac{y^{n}}{n!}\leq \frac{y^{k}}{k!}\exp(y),
\end{align}
and setting $y=M\tau\Lambda$ we can get a bound on the infinite sum in (\ref{eq3-21}):
\begin{align}
   \label{eq3-23}
   \dnorm{T_{\tau}-S_{1}^{(ran)}(\tau) }&\leq 2\frac{M^{3}\tau^{3}\Lambda^{3}}{3!}\exp(M\tau\Lambda).
\end{align}
Replacing $\tau$ with $t/N$ in (\ref{eq3-23}) and substituting into (\ref{eq3-7a}) yields the bound
\begin{align}
   \label{eq3-24}
   \dnorm{T_{t}-S_{1}^{(ran)}\left(\frac{t}{N}\right)^{N}}&\leq  \frac{M^{3}t^{3}\Lambda^{3}}{3N^{2}}\exp(\frac{Mt\Lambda}{N}).
\end{align}
Since $N$ is chosen such that $Mt\Lambda/N \leq 1$, we have $\exp(Mt\Lambda/N) \leq \exp(1) = e$, which gives the bound
\begin{align}
   \label{eq3-25}
   \dnorm{T_{t}-S_{1}^{(ran)}\left(\frac{t}{N}\right)^{N}}&\leq e\frac{M^{3}t^{3}\Lambda^{3}}{3N^{2}}.
\end{align}
Then letting
\begin{align}
   \label{eq3-26}
   \epsilon \geq e\frac{M^{3}t^{3}\Lambda^{3}}{3N^{2}},
\end{align}
we have shown that,
\begin{align}
   \dnorm{T_{t}-S_{1}^{(ran)}\left(\frac{t}{N}\right)^{N}}&\leq \epsilon.
\end{align}
From (\ref{eq3-26}) we find the bound on $N$ to be
\begin{align}
   \label{eq3-27}
   N \geq e^{1/2}\frac{M^{3/2}t^{3/2}\Lambda^{3/2}}{\sqrt{3\epsilon}}
\end{align}
and this completes the proof.
\end{proof}

\section{Second Order Randomised Product Formula}
\label{section4}
The second order randomised Trotter-Suzuki product formula is not much more complicated than its deterministic counterpart. It is constructed by considering a convex sum of all permutations of the exponentials in the second order Trotter-Suzuki product formula. More precisely, consider the symmetric group $\mathrm{Sym}(M)$ which is the group of all permutations of the elements of the set $\{1,...,M\}$. For any permutation $\sigma \in \mathrm{Sym}(M)$ we define,
\begin{align}
   \label{eq4-1}
   S_{2}^{\sigma}(\tau):=\prod_{j=1}^{M}e^{\frac{\tau}{2}\Lk{\sigma(j)}}\prod_{k=M}^{1}e^{\frac{\tau}{2}\Lk{\sigma(k)}},
\end{align}
which is a second order Trotter-Suzuki product formula whose exponentials are permuted by the permutation $\sigma \in \mathrm{Sym}(M)$. Using this, we can construct the approximation to the total channel $T_{\tau}$ by taking a convex combination of $S_{2}^{\sigma}$ for all $\sigma \in \mathrm{Sym}(M)$,
\begin{align}
   \label{eq4-2}
   S_{2}^{(ran)}(\tau)=\frac{1}{M!}\sum_{\sigma \in \mathrm{Sym}(M)} S_{2}^{\sigma}(\tau),
\end{align}
where the $1/M!$ is present because $|\mathrm{Sym}(M)|=M!$. The following theorem shows the error bound and the bound on $N$ for the second order randomised Trotter-Suzuki product formula. The proof of this theorem is done in a similar way to the proof of the randomised product formulas in \cite{childs2019faster}. However, it also relies on the mixing lemma \cite{campbell2017shorter,hastings2016turning}, which as stated before is not applicable to open quantum systems, therefore we give a more direct proof.

\begin{theorem}
\label{thoerem4}
   Given the generator $\mathcal{L}$ as in equation (\ref{eq10}) of a quantum channel $T_{t}$ and some time $t\geq 0$. Define the second order randomised product formula as in equation (\ref{eq4-2}). Let  $\Lambda:=\max_{k}\dnorm{\Lk{k}}$ and choose $N$ such that $Mt\Lambda/N \leq 1$. Then,
   \begin{align}
       \label{eq4-3}
       \dnorm{T_{t}-S_{2}^{(ran)}\left(\frac{t}{N}\right)^{N}}\leq \frac{e(2\Lambda t)^{3}M^{2}}{N^{2}},
   \end{align}
   where
   \begin{align}
       \label{eq4-4}
       \epsilon\geq \frac{e(2\Lambda t)^{3}M^{2}}{N^{2}},
   \end{align}
   and
   \begin{align}
       \label{eq4-5}
       N\geq \frac{e^{1/2}(2 \Lambda t)^{3/2}M}{\epsilon^{1/2}}.
   \end{align}
\end{theorem}

\noindent In proving the following theorem we will need to Taylor expand the formula $S_{2}^{(ran)}$. However this may be a complicated and challenging task to do directly. Instead, we want to consider what an arbitrary order term looks like in the expansion.  The following lemma, which we shall call the randomisation lemma, will tell us what the $s$-th order non-degenerate term looks like in the expansion of $S_{2}^{(ran)}$, where $0 \leq s \leq M.$ We use the word non-degenerate to describe a product of constituent generators $\Lk{k}$ which is pairwise different. To make out calculations easier we rewrite the second order randomised formula $S_{2}^{(ran)}$ in the following general way
\begin{align}
   \label{eq4-6}
   S_{2}^{(ran)}(\tau)=\frac{1}{M!}\sum_{\sigma\in \mathrm{Sym}(M)}&\exp(q_{1}\tau\Lk{\sigma(\pi_{1}(1))})...\exp(q_{1}\tau\Lk{\sigma(\pi_{1}(M))})\times \nonumber\\
   &\exp(q_{2}\tau\Lk{\sigma(\pi_{2}(1))})...\exp(q_{2}\tau\Lk{\sigma(\pi_{2}(M))}),
\end{align}
where $\tau\geq 0$, and $q_{1},q_{2}\in \mathbb{R}^{+}$ that we will define at a later stage and $\pi_{1},\pi_{2} \in \mathrm{Sym}(M)$ such that $\pi_{1}=\mathrm{id}$ and $\pi_{2}$ is defined as,
\begin{align}
   \label{eq4-7}
   \pi_{2}=\begin{pmatrix}
       1 & 2 & ... & M\\
       M & M-1 & ... & 1
   \end{pmatrix}.
\end{align}

\noindent We now state the randomisation lemma.
\begin{lemma}
\label{lemma2}
   Given the second order randomised product formula as in equation (\ref{eq4-6}), let $s\in \mathbb{N}$ such that $0 \leq s \leq M$. The $s$-th order non-degenerate term of $S_{2}^{(ran)}$ is
   \begin{align}
       \label{eq4-8}
       \frac{\tau^{s}}{s!}\sum_{\substack{m_{1},...,m_{s}=1 \\ \text{pairwise different}}}^{M}\Lk{m_{1}} ... \Lk{m_{s}}.
   \end{align}
\end{lemma}
\begin{proof}
   We start by expanding each exponential in (\ref{eq4-6}) in a Taylor series and
we take all possible products of $s$ terms from each of the Taylor expansions. We observe that in (\ref{eq4-6}) the exponentials are arranged in an array of two rows and $M$ columns. Using the indicies $\kappa_{1},...,\kappa_{s}$ and $l_{1},...,l_{s}$ to label the rows and columns, respectively, of the exponential from which the terms are chosen. To avoid double counting, we ensure that $\kappa_{1}\leq \kappa_{2} \leq ... \leq \kappa_{s}$. Within each row, we also want to ensure that we have smaller column indicies first. Since $\pi_{1}$ and $\pi_{2}$ are bijective to get the non-degenerate term, we require that $l_{1},...l_{s}$ are pairwise different. The $s$-th order non-degenerate term is,
\begin{align}
   \label{eq4-10}
   \frac{1}{M!}\sum_{\sigma\in \mathrm{Sym}(M)} \ \sum_{\substack{\kappa_{1},...,\kappa_{s}=1\\\kappa_{1}\leq ...\leq \kappa_{s}}}^{2} \ \sum_{\substack{\pi_{\kappa_{1}(l_{1}),...,\pi_{\kappa_{s}}(l_{s})=1}\\\text{pairwise different}}}^{M}\left(q_{\kappa_{1}}\tau \Lk{\sigma(\pi_{\kappa_{1}}(l_{1}))}\right)...\left(q_{\kappa_{s}}\tau \Lk{\sigma(\pi_{\kappa_{s}}(l_{s}))}\right).
\end{align}
A direct calculation shows that,
\begin{align}
   \label{eq4-11}
   &\frac{1}{M!}\sum_{\sigma\in \mathrm{Sym}(M)} \ \sum_{\substack{\kappa_{1},...,\kappa_{s}=1\\\kappa_{1}\leq ...\leq \kappa_{s}}}^{2} \ \sum_{\substack{\pi_{\kappa_{1}(l_{1}),...,\pi_{\kappa_{s}}(l_{s})=1}\\\text{pairwise different}}}^{M}\left(q_{\kappa_{1}}\tau \Lk{\sigma(\pi_{\kappa_{1}}(l_{1}))}\right)...\left(q_{\kappa_{s}}\tau \Lk{\sigma(\pi_{\kappa_{s}}(l_{s}))}\right)\nonumber\\
   &=\frac{1}{M!}\sum_{\sigma\in \mathrm{Sym}(M)} \ \sum_{\substack{\kappa_{1},...,\kappa_{s}=1\\\kappa_{1}\leq ...\leq \kappa_{s}}}^{2} \ \sum_{\substack{\pi_{\kappa_{1}(l_{1}),...,\pi_{\kappa_{s}}(l_{s})=1}\\\text{pairwise different}}}^{M} \ \sum_{\substack{m_{1}=\sigma(\pi_{\kappa_{1}}(l_{1})),..., \\ m_{s}=\sigma(\pi_{\kappa_{s}}(l_{s}))}}\left( q_{\kappa_{1}}\tau\Lk{m_{1}}\right)...\left( q_{\kappa_{s}}\tau\Lk{m_{s}}\right)\nonumber\\
   &=\frac{1}{M!}\sum_{\substack{m_{1},...,m_{s}=1\\\text{pairwise different}}}^{M} \ \sum_{\substack{\kappa_{1},...,\kappa_{s}=1\\\kappa_{1}\leq ...\leq \kappa_{s}}}^{2} \ \sum_{\substack{\pi_{\kappa_{1}(l_{1}),...,\pi_{\kappa_{s}}(l_{s})=1}\\\text{pairwise different}}}^{M}\sum_{\substack{\sigma \in \mathrm{Sym}(M)\\m_{1}=\sigma(\pi_{\kappa_{1}}(l_{1})),..., \\ m_{s}=\sigma(\pi_{\kappa_{s}}(l_{s}))}}\left( q_{\kappa_{1}}\tau\Lk{m_{1}}\right)...\left( q_{\kappa_{s}}\tau\Lk{m_{s}}\right).
\end{align}
The last sum in equation (\ref{eq4-11}) is a permutation of all pairwise different $m_{1},...,m_{s}$ and we observe that for a fixed $m_{1},...,m_{s}$ there are $(M-s)!$ ways we can permute the rest of the indicies so that $m_{1},...,m_{s}$ is unchanged. Therefore, we remove this sum and add a factor $(M-s)!$, leading to the following expression for the $s$-th order non-degenerate term:
\begin{align}
   \label{eq4-12}
   \frac{(M-s)!}{M!}\sum_{\substack{m_{1},...,m_{s}=1\\\text{pairwise different}}}^{M} \ \left[ \sum_{\substack{\kappa_{1},...,\kappa_{s}=1\\\kappa_{1}\leq ...\leq \kappa_{s}}}^{2} \ \sum_{\substack{\pi_{\kappa_{1}(l_{1}),...,\pi_{\kappa_{s}}(l_{s})=1}\\\text{pairwise different}}}^{M} (q_{\kappa_{1}}\tau)...(q_{\kappa_{s}}\tau)\right]\Lk{m_{1}}...\Lk{m_{s}}.
\end{align}
Now we need to calculate the sum in the brackets in equation (\ref{eq4-12}). This sum depends solely on the row indicies, so by letting $r_{1}$ and $r_{2}$ be the number of terms picked from row one and row two respectively, we can express the summand as
\begin{align}
   \label{eq4-13}
   (q_{1}\tau)^{r_{1}}(q_{2}\tau)^{r_{2}}.
\end{align}
All that remains is to determine the value of the sums which can be found using combinatorial arguments. The number of ways we can choose $l_{1},...,l_{s}$ pairwise different is given by
\begin{align}
   \label{eq4-14}
   M(M-1)...(M-(s+1))=\frac{M!}{(M-s)!}.
\end{align}
However, when we apply the permutations $\pi_{1}$ and $\pi_{2}$ we may double count some terms. In particular if $\kappa_{i}=\kappa_{i+1}$, we have to pick terms from the row $\kappa_{i}$ and we must have $l_{i} <l_{i+1}$. This implies that the ordering of $\pi_{\kappa_{i}}(l_{i})$ and $\pi_{\kappa_{i+1}}(l_{i+1})$ is uniquely determined. Altogether, we have then overcounted by a factor of $r_{1}!r_{2}!$. Therefore, we have
\begin{align}
   \label{eq4-15}
   \sum_{\substack{\kappa_{1},...,\kappa_{s}=1\\\kappa_{1}\leq ...\leq \kappa_{s}}}^{2} \ \sum_{\substack{\pi_{\kappa_{1}(l_{1}),...,\pi_{\kappa_{s}}(l_{s})=1}\\\text{pairwise different}}}^{M} (q_{\kappa_{1}}\tau)...(q_{\kappa_{s}}\tau)&=\sum_{\substack{r_{1},r_{2}=0\\r_{1}+r_{2}=s}}^{s}\frac{M!}{(M-s)!}\frac{(q_{1}\tau)^{r_{1}}(q_{2}\tau)^{r_{2}}}{r_{1}!r_{2}!}\nonumber\\
   &=\frac{M!}{(M-s)!}\frac{[(q_{1}+q_{2})\tau]^{s}}{s!},
\end{align}
where the last equality is a result of the multinomial theorem. Substituting (\ref{eq4-15}) into (\ref{eq4-12}) we get the $s$-th order non-degenerate term
\begin{align}
   \label{eq4-16}
   \frac{(M-s)!}{M!}\sum_{\substack{m_{1},...,m_{s}=1\\\text{pairwise different}}}^{M} \ \frac{M!}{(M-s)!}\frac{[(q_{1}+q_{2})\tau]^{s}}{s!}\Lk{m_{1}}...\Lk{m_{s}}.
\end{align}
Simplifying this expression yields
\begin{align}
   \label{eq4-17}
   \frac{[(q_{1}+q_{2})\tau]^{s}}{s!}\sum_{\substack{m_{1},...,m_{s}=1\\\text{pairwise different}}}^{M}\Lk{m_{1}}...\Lk{m_{s}}.
\end{align}
Since $S_{2}^{(ran)}(\tau)$ is atleast first order accurate it implies that for $s=1$ this term should cancel exactly with the first order term in the Taylor expansion of $T_{\tau}$. This implies that $q_{1}+q_{2}=1$, allowing us to set $q_{1}=q_{2}=1/2$ as in the definition of $S_{2}^{(ran)}$ in equation (\ref{eq4-1}). This leads to the desired expression for the $s$-th order non-degenerate term:
\begin{align}
   \label{eq4-18}
   \frac{\tau^{s}}{s!}\sum_{\substack{m_{1},...,m_{s}=1\\\text{pairwise different}}}^{M}\Lk{m_{1}}...\Lk{m_{s}}.
\end{align}
\end{proof}
\noindent Since we want to compute the error between our second order randomised formula and the channel $T_{\tau}$, it will be useful to understand the form of the $s$-th order non-degenerate term of $T_{\tau}$. The following lemma gives the $s$-th order non-degenerate term of $T_{\tau}$.
\begin{lemma}
\label{lemma3}
   The $s$-th order non-degenerate term of $T_{\tau}$ is
   \begin{align}
       \label{eq4-19}
       \frac{\tau^{s}}{s!}\sum_{\substack{m_{1},...,m_{s}=1\\\text{pairwise different}}}^{M}\Lk{m_{1}}...\Lk{m_{s}}.
   \end{align}
\end{lemma}

\begin{proof}
   Consider the Taylor expansion of $T_{\tau}$
   \begin{align}
       \label{4-20}
       T_{\tau}=\sum_{s=0}^{\infty}\frac{\tau^{s}}{s!}\mathcal{L}^{s}=\sum_{s=0}^{\infty}\frac{\tau^{s}}{s!}\left(\sum_{k=1}^{M}\Lk{k}\right)^{s},
   \end{align}
   we can write $\left(\sum_{k=1}^{M}\Lk{k}\right)^{s}$ as
   \begin{align}
       \label{eq4-21}
       \left(\sum_{k=1}^{M}\Lk{k}\right)^{s}=\sum_{m_{1},...,m_{s}=1}^{M}\Lk{m_{1}}...\Lk{m_{s}}.
   \end{align}
   To obtain the non-degenerate term, we require that $m_{1},...,m_{s}$ be pairwise different so the non-degenerate term is
   \begin{align}
       \label{eq4-22}
       \frac{\tau^{s}}{s!}\sum_{\substack{m_{1},...,m_{s}=1\\ \text{pairwise different}}}^{M}\Lk{m_{1}}...\Lk{m_{s}}.
   \end{align}
\end{proof}

\noindent Noting that the arbitrary $s$-th order term has both a degenerate and non-degenerate part, we then aim to obtain a bound on the norm of the $s$-th order degenerate terms of $T_{\tau}$ and $S_{2}^{(ran)}$ so that we can bound the error. The lemma below gives the bound on the norm of the $s$-th order degenerate term.
\begin{lemma}
\label{lemma4}
   Let $\mathcal{L}$ be defined as in equation (\ref{eq10}) and let $\Lambda:=\max_{k}\dnorm{\Lk{k}}$. Define the ideal evolution for a small time step $\tau\geq 0$ as $T_{\tau}=\exp(\tau\mathcal{L})$ and define the second order randomised formula $S_{2}^{(ran)}$ as in equation (\ref{eq4-2}) and let $s$ be a natural number such that $0\leq s \leq M$. The norm of the $s$-th order degenerate term of the ideal evolution $T_{\tau}$ is at most
   \begin{align}
       \label{eq4-23}
       \frac{(\tau\Lambda)^{s}}{s!}[M^{s}-M(M-1)...(M-(s+1))].
   \end{align}
   The norm of the $s$-th order degenerate term of $S_{2}^{(ran)}(\tau)$ is at most,
   \begin{align}
       \label{eq4-24}
       \frac{(\tau \Lambda)^{s}}{s!}[M^{s}-M(M-1)...(M-(s+1))].
   \end{align}
\end{lemma}
\begin{proof}
To derive the bound on the norm of the s-th order degenerate term, we use a combinatorial argument. The complete set of $s$-th order terms can be partitioned into two disjoint sets: those with pairwise different indices (non-degenerate) and those with at least one repeated index (degenerate). The bound on the norm of the entire s-th order term is the sum of the norm bounds of all constituent terms. Our approach is to subtract the contribution of the non-degenerate terms from this total bound. The bound for the non-degenerate part is obtained by counting only the terms with distinct indices. The remaining contribution must therefore come from the degenerate terms. This counting-based approach is valid and is used in related literature on randomized simulation \cite{childs2019faster}. 

We first consider the degenerate term of the ideal evolution $T_{\tau}$. The upper bound on the norm of the degenerate term is given by,
\begin{align}
    \dnorm{\frac{\tau^{s}}{s!}\sum_{\substack{m_{1},...,m_{s}=1\\ \text{degenerate}}}^{M}\Lk{m_{1}}...\Lk{m_{s}}} &\leq \frac{\tau^{s}}{s!}\sum_{\substack{m_{1},...,m_{s}=1\\ \text{degenerate}}}^{M}\dnorm{\Lk{m_{1}}}...\dnorm{\Lk{m_{s}}}\\
    &\leq \frac{\tau^{s}\Lambda^{s}}{s!}\sum_{\substack{m_{1},...,m_{s}=1\\ \text{degenerate}}}^{M},\\
    &=\frac{\tau^{s}\Lambda^{s}}{s!}\left[\sum_{m_{1},...,m_{s}=1}^{M}-\sum_{\substack{m_{1},...,m_{s}=1\\ \text{non-degenerate}}}^{M}\right],\\
    &= \frac{\tau^{s}\Lambda^{s}}{s!}\left[M^{s}-M(M-1)...(M-(s+1))\right]
\end{align}
where we use the fact that the number of degenerate terms in the sum is equal to the total number of terms minus the number of non-degenerate terms. We prove in a similar way that the upper bound on the norm of the degenerate term of $S_{2}^{(\mathrm{ran})}$.

\end{proof}

\noindent Now we wish to bound the error for a fixed order term in the difference between the ideal evolution and the second order randomised formula. The lemma below will show this error bound.

\begin{lemma}
   \label{lemma5}   
   Given a generator $\mathcal{L}$ as in equation (\ref{eq10}) and the ideal evolution $T_{\tau}$ for some small time step $\tau\geq 0$ as well as the second order randomised formula $S_{2}^{(ran)}(\tau)$ defined in equation (\ref{eq4-2}). Let the error superoperator be defined as $E(\tau)=T_{\tau}-S_{2}^{(ran)}(\tau)$ and let $E_{s}(\tau)$ be the $s$-th order term in the Taylor expansion of $E(\tau)$. Then, the diamond norm of $E_{s}(\tau)$ is bounded for each order $s$ as,
   \begin{align}
       \label{eq4-34}
       \dnorm{E_{s}(\tau)}\leq 
       \begin{cases}
       0, \text{ for }0\leq s\leq 2,\\
       \\
       \frac{(\Lambda \tau)^{s} M^{s-1} }{(s-2)!}, \text{ for } s> 2.
   \end{cases}
   \end{align}
\end{lemma}

\begin{proof}
   For $s\leq 2$, the formula $S_{2}^{\sigma}(\tau)$ is exact so it cancels with all the second order terms and since we sum convexly in the second order randomised formula, all second order terms in
   \begin{align}
       \label{eq4-35}
       \frac{1}{M!}\sum_{\sigma \in \mathrm{Sym}(M)}S_{2}^{\sigma}(\tau),
   \end{align}
   cancel with all second order terms in $T_{\tau}$. Therefore the error is zero for $s\leq 2$.
   For $s>2$ we need to bound the error in of the $s$-th order term in the difference $T_{\tau}-S_{2}^{(ran)}(\tau)$. We know from Lemma \ref{lemma2}. and Lemma \ref{lemma3}. that the $s$-th order non-degenerate terms of $T_{\tau}$ and $S_{2}^{(ran)}(\tau)$ are the same and they will cancel when we take the difference $T_{\tau}-S_{2}^{(ran)}(\tau)$. This means that the only terms we need to consider are the $s$-th order degenerate terms. From Lemma \ref{lemma4}, we see that the bound on the error of the difference is the sum of the bounds of the degenerate terms. That is
   \begin{align}
   \label{eq4-36}
       2\frac{(\tau \Lambda)^{s}}{s!}[M^{s}-M(M-1)...(M-(s+1))].
   \end{align}
   Now we need to obtain a bound for the factor $[M^{s}-M(M-1)...(M-(s+1))]$. Using \cite{childs2019faster} we can obtain the following bound
   \begin{align}
       \label{eq4-37}
       [M^{s}-M(M-1)...(M-(s+1))]\leq \begin{pmatrix}
           s\\
           2
       \end{pmatrix}M^{s-1}=\frac{s!}{2!(s-2)!}M^{s-1}.
   \end{align}
   This leads us to the error bound for $s>2$:
   \begin{align}
       \label{eq4-38}
       2\frac{(\tau \Lambda)^{s}}{s!}[M^{s}-M(M-1)...(M-(s+1))]\leq 2\frac{(\tau \Lambda)^{s}}{s!}\frac{s!}{2!(s-2)!}M^{s-1}=\frac{(\tau \Lambda)^{s}M^{s-1}}{(s-2)!},
   \end{align}
   which is the desired bound.
\end{proof}

\noindent We are now able to prove Theorem \ref{thoerem4} and obtain the bound on the precision $\epsilon$ and number of steps $N$ for the second order randomised formula $S_{2}^{(ran)}$.

\begin{proof}{(of Theorem \ref{thoerem4}.)} We start by applying Lemma \ref{lemma1}, to (\ref{eq4-3}) which yields
\begin{align}
   \label{eq4-39}
   \dnorm{T_{t}-S_{2}^{(ran)}\left(\frac{t}{N}\right)^{N}}\leq N\dnorm{T_{\tau}-S_{2}^{(ran)}(\tau)}.
\end{align}
   From Lemma \ref{lemma5}, we have that the bound on $\dnorm{T_{\tau}-S_{2}^{(ran)}(\tau)}$ is, for $s=3$,
   \begin{align}
       \label{eq4-40}
       \dnorm{T_{\tau}-S_{2}^{(ran)}(\tau)}\leq \sum_{s=3}^{\infty}\frac{(\Lambda \tau)^{s} M^{s-1} }{(s-2)!}
   \end{align}
   Lemma F.2 from the supplementary information of \cite{childs2018toward} states that for some $y \geq 0 \in \mathbb{R}$ and $k \in \mathbb{N}$
\begin{align}
   \label{eq4-41}
   \sum_{n=k}^{\infty}\frac{y^n}{n!}\leq \frac{y^{k}}{k!}\exp(y).
\end{align}
We can use this lemma to bound the sum in equation (\ref{eq4-40}) by setting $y=\Lambda \tau M$ and $k=3$,
\begin{align}
   \label{eq4-42}
   \dnorm{T_{\tau}-S_{2}^{(ran)}(\tau)}\leq \frac{(\Lambda \tau)^{3}M^{2}}{(3-2)!}\exp(\Lambda \tau M).
\end{align}
Using the fact that $\tau=t/N$, we have
\begin{align}
\label{eq4-43}
   \dnorm{T_{\tau}-S_{2}^{(ran)}(\tau)}\leq \frac{(\Lambda t)^{3}M^{2}}{N^{3}}\exp(\Lambda M t/N).
\end{align}
Since $N$ is chosen such that $Mt\Lambda/N \leq 1$, we have $\exp(\Lambda M t/N) \leq \exp(1) = e$. Hence the bound is
\begin{align}
   \label{eq4-44}
   \dnorm{T_{\tau}-S_{2}^{(ran)}(\tau)}\leq e\frac{(\Lambda t)^{3}M^{2}}{N^{3}}.
\end{align}
Substituting this into (\ref{eq4-39}) yields,
\begin{align}
   \label{eq4-45}
   \dnorm{T_{t}-S_{2}^{(ran)}\left(\frac{t}{N}\right)^{N}}\leq N e\frac{(\Lambda t)^{3}M^{2}}{N^{3}}=e\frac{(\Lambda t)^{3}M^{2}}{N^{2}}.
\end{align}
Now, let $\epsilon \geq 0$ such that
\begin{align}
   \label{eq4-46}
   \epsilon \geq e\frac{(\Lambda t)^{3}M^{2}}{N^{2}}
\end{align}
which leads to the bound
\begin{align}
   \label{eq4-47}
   N\geq e^{1/2}\frac{(\Lambda t)^{3/2}M}{\epsilon^{1/2}}.
\end{align}

\end{proof}
\section{The QDRIFT Channel For Simulating OQS}
\label{section5}
In this section we will outline how we can use the QDRIFT channel \cite{campbell2019random} to simulate Markovian OQS. Consider the generator $\mathcal{L}$ in equation (\ref{eq8}), this form shall be used throughout the rest of this section. We start by defining the quantity $\Gamma$, which is the sum of all the decay rates in $\mathcal{L}$. That is
\begin{align}
   \label{eq5-1}
   \Gamma=\sum_{k=1}^{M}\gamma_{k}.
\end{align}
To define the QDRIFT channel we must define the small time step $\omega=\frac{t \Gamma}{N}$. The QDRIFT channel, probabilistically implements a constituent channel $e^{\omega\mathcal{L}_{k}}$ with some probability $p_{k}$ which depends on the decay rate $\gamma_{k}$ in the generator in the following way,
\begin{align}
   \label{eq5-2}
   p_{k}=\frac{\gamma_{k}}{\Gamma}.
\end{align}

\noindent It is evident from the definition of $p_{k}$ that $\sum_{k=1}^{M}p_{k}=1$ and that for larger $\gamma_{k}$, the QDRIFT channel is more likely to apply $e^{\omega\mathcal{L}_{k}}$. While this process is random, the probabilities $p_{k}$ have a bias built into them so that with many repetitions the evolution stochastically 'drifts' towards the ideal evolution $T_{t}$. Since each constituent channel is sampled independently, the process is entirely Markovian and we can consider the evolution resulting from a single random operation. The QDRIFT channel, which shall be denoted by $\mathcal{E}^{(QD)}_{\omega}$ has the form,
\begin{align}
   \label{eq5-3}
   \mathcal{E}^{(QD)}_{\omega}(\rho)=\sum_{k=1}^{M}p_{k}e^{\omega\mathcal{L}_{k}}.
\end{align}

The error analysis below is controlled by the parameter $\Omega := \max_{k}\dnorm{\mathcal{L}_{k}}$. As with $\Lambda$, we stress that the $\mathcal{L}_{k}$ are the rate-normalised \emph{generators} of equation (\ref{eq9}) and not quantum channels; they are Hermiticity-preserving but not trace-preserving, so their diamond norms are in general not equal to one. The distinction between $\Lambda$ and $\Omega$ is that $\Lambda=\max_{k}\dnorm{\hat{\mathcal{L}}_{k}}$ is taken over the rate-absorbed generators $\hat{\mathcal{L}}_{k}=\gamma_{k}\mathcal{L}_{k}$, so that $\Lambda$ captures both the decay rate $\gamma_{k}$ and the operator magnitude, whereas $\Omega$ is the rate-normalised version, with the rates instead collected into $\Gamma$ through the sampling probabilities $p_{k}=\gamma_{k}/\Gamma$. The CPTP maps appearing in the circuit are the constituent exponentials $e^{\omega\mathcal{L}_{k}}$, which have unit diamond norm.

\noindent We now state the following theorem which outlines how the QDRFIT channel approximates $T_{t}$. We shall save the discussion of how to implement this QDRIFT channel for a later section.

\begin{theorem}
   Given the generator $\mathcal{L}$ as in equation (\ref{eq8}) of a quantum channel $T_{t}$ and some time $t\geq 0$. Let $\Omega := \max_{k}\dnorm{\mathcal{L}_{k}}$ and choose $N$ such that $t\Gamma \Omega/N\leq 1$.  Then,
   \begin{align}
       \label{eq5-4}
       \dnorm{T_{t}-\left(\mathcal{E}^{(QD)}_{\omega} \right)^{N}}=\dnorm{T_{t}-\left(\sum_{k=1}^{M}p_{k}e^{\omega\mathcal{L}_{k}}\right)^{N}}\leq \frac{et^{2}\Gamma^{2}\Omega^{2}}{N}
   \end{align}
   where
   \begin{align}
       \label{eq5-5}
       \epsilon \geq \frac{et^{2}\Gamma^{2}\Omega^{2}}{N}
   \end{align}
   and,
   \begin{align}
       \label{eq5-6}
       N \geq \frac{et^{2}\Gamma^{2}\Omega^{2}}{\epsilon}.
   \end{align}
\end{theorem}

\begin{proof}
   Using Lemma \ref{lemma1}. we see that,
   \begin{align}
   \label{eq5-7}
       \dnorm{T_{t}-\left(\mathcal{E}^{(QD)}_{\omega} \right)^{N}}&=\dnorm{\exp(t\mathcal{L})-\left(\mathcal{E}^{(QD)}_{\omega} \right)^{N}},\nonumber\\
       \nonumber\\
       &=\dnorm{\exp(\frac{t}{N}\sum_{k=1}^{M}\gamma_{k}\mathcal{L}_{k})^{N}-\left(\sum_{k=1}^{M}p_{k}\exp(\omega\mathcal{L}_{k}) \right)^{N}}\nonumber\\
       \nonumber\\
       &\leq N\dnorm{\exp(\frac{t}{N}\sum_{k=1}^{M}\gamma_{k}\mathcal{L}_{k})-\left(\sum_{k=1}^{M}p_{k}\exp(\omega\mathcal{L}_{k}) \right)}.
   \end{align}
   
\noindent To complete the proof, we need to find a bound on $\dnorm{\exp(\frac{t}{N}\sum_{k=1}^{M}\gamma_{k}\mathcal{L}_{k})-\left(\sum_{k=1}^{M}p_{k}\exp(\omega\mathcal{L}_{k}) \right)}$. We start by expanding $\exp(\frac{t}{N}\sum_{k=1}^{M}\gamma_{k}\mathcal{L}_{k})$ in a Taylor series as
\begin{align}
   \label{eq5-8}
   \exp(\frac{t}{N}\sum_{k=1}^{M}\gamma_{k}\mathcal{L}_{k})&=\mathbb{1}+\frac{t}{N}\sum_{k=1}^{M}\gamma_{k}\mathcal{L}_{k}+\sum_{\nu=2}^{\infty}\frac{(t/N)^{\nu}}{\nu!}\left(\sum_{k=1}^{M}\gamma_{k}\mathcal{L}_{k}\right)^{\nu}\nonumber\\
   &=\mathbb{1}+\frac{t}{N}\mathcal{L}+\sum_{\nu=2}^{\infty}\frac{(t/N)^{\nu}}{\nu!}\mathcal{L}^{\nu}.
\end{align}
We can also expand the exponential in the term $\sum_{k=1}^{M}p_{k}\exp(\omega\mathcal{L}_{k})$ in a Taylor series as follows,
\begin{align}
   \label{eq5-9}
   \sum_{k=1}^{M}p_{k}\exp(\omega\mathcal{L}_{k})&=\sum_{k=1}^{M}p_{k}\left(\sum_{\nu=0}^{\infty}\frac{\omega^{\nu}}{\nu!}\mathcal{L}_{k}^{\nu}\right),\nonumber\\
   \nonumber\\
   &=\sum_{k=1}^{M}p_{k}\left(\mathbb{1}+\omega\mathcal{L}_{k}+\sum_{\nu=2}^{\infty}\frac{\omega^{\nu}}{\nu!}\mathcal{L}_{k}^{\nu}\right),\nonumber\\
   \nonumber\\
   &=\mathbb{1}+\sum_{k=1}^{M}p_{k}\omega\mathcal{L}_{k}+\sum_{k=1}^{M}p_{k}\sum_{\nu=2}^{\infty}\frac{\omega^{\nu}}{\nu!}\mathcal{L}_{k}^{\nu},\nonumber\\
   \nonumber\\
   &=\mathbb{1}+\sum_{k=1}^{M}\frac{\gamma_{k}}{\Gamma}\omega\mathcal{L}_{k}+\sum_{k=1}^{M}\frac{\gamma_{k}}{\Gamma}\sum_{\nu=2}^{\infty}\frac{\omega^{\nu}}{\nu!}\mathcal{L}_{k}^{\nu},\nonumber\\
   \nonumber\\
   &=\mathbb{1}+\frac{\omega}{\Gamma}\mathcal{L}+\sum_{k=1}^{M}\frac{\gamma_{k}}{\Gamma}\sum_{\nu=2}^{\infty}\frac{\omega^{\nu}}{\nu!}\mathcal{L}_{k}^{\nu}.
\end{align}
We can now compute the norm $\dnorm{\exp(\frac{t}{N}\sum_{k=1}^{M}\gamma_{k}\mathcal{L}_{k})-\left(\sum_{k=1}^{M}p_{k}\exp(\omega\mathcal{L}_{k}) \right)}$ as follows: if we use the fact that $\omega=t\Gamma/N$ then we see that the zeroth and first order terms in (\ref{eq5-8}) and (\ref{eq5-9}) cancel leaving us with
\begin{align}
\label{eq5-10}
   \dnorm{\exp(\frac{t}{N}\sum_{k=1}^{M}\gamma_{k}\mathcal{L}_{k})-\left(\sum_{k=1}^{M}p_{k}\exp(\omega\mathcal{L}_{k}) \right)}=\dnorm{\sum_{\nu=2}^{\infty}\frac{t^{\nu}}{N^{\nu}\nu!}\mathcal{L}^{\nu}-\sum_{k=1}^{M}\frac{\gamma_{k}}{\Gamma}\sum_{\nu=2}^{\infty}\frac{\omega^{\nu}}{\nu!}\mathcal{L}_{k}^{\nu}}.
\end{align}
Using the sub-additive and sub-multiplicative properties of the diamond norm, one obtains
\begin{align}
   \label{eq5-11}
   \dnorm{\exp(\frac{t}{N}\sum_{k=1}^{M}\gamma_{k}\mathcal{L}_{k})-\left(\sum_{k=1}^{M}p_{k}\exp(\omega\mathcal{L}_{k}) \right)}\leq \sum_{\nu=2}^{\infty}\frac{t^{\nu}}{N^{\nu}\nu!}\dnorm{\mathcal{L}}^{\nu}+\sum_{k=1}^{M}\frac{\gamma_{k}}{\Gamma}\sum_{\nu=2}^{\infty}\frac{\omega^{\nu}}{\nu!}\dnorm{\mathcal{L}_{k}}^{\nu}.
\end{align}
At this point we wish to find bounds on the diamond norm of the generator $\mathcal{L}$ and the superoperators $\mathcal{L}_{k}$. By definition we have that $\dnorm{\mathcal{L}_{k}}\leq \Omega$, which yields
\begin{align}
   \label{eq5-12}
    \dnorm{\mathcal{L}_{k}}^{\nu}\leq \Omega^{\nu}.
\end{align}
For the generator $\mathcal{L}$ we have
\begin{align}
   \label{eq5-13}
   \dnorm{\mathcal{L}}=\dnorm{\sum_{k=1}^{M}\gamma_{k}\mathcal{L}_{k}}\leq \sum_{k=1}^{M}\gamma_{k}\dnorm{\mathcal{L}_{k}}\leq \sum_{k=1}^{M}\gamma_{k}\Omega=\Gamma \Omega
\end{align}
which implies that
\begin{align}
   \label{eq5-14}
   \dnorm{\mathcal{L}}^{\nu}\leq \Gamma^{\nu}\Omega^{\nu}.
\end{align}
Substituting (\ref{eq5-12}) and (\ref{eq5-14}) into (\ref{eq5-11}) produces
\begin{align}
   \label{eq5-15}
   \dnorm{\exp(\frac{t}{N}\sum_{k=1}^{M}\gamma_{k}\mathcal{L}_{k})-\left(\sum_{k=1}^{M}p_{k}\exp(\omega\mathcal{L}_{k}) \right)}&\leq \sum_{\nu=2}^{\infty}\frac{t^{\nu}\Gamma^{\nu}\Omega^{\nu}}{N^{\nu}\nu!}+\sum_{k=1}^{M}\frac{\gamma_{k}}{\Gamma}\sum_{\nu=2}^{\infty}\frac{\omega^{\nu}}{\nu!}\Omega^{\nu},\nonumber\\
   \nonumber\\
   &=\sum_{\nu=2}^{\infty}\frac{t^{\nu}\Gamma^{\nu}\Omega^{\nu}}{N^{\nu}\nu!}+\frac{\Gamma}{\Gamma}\sum_{\nu=2}^{\infty}\frac{t^{\nu}\Gamma^{\nu}\Omega^{\nu}}{N^{\nu}\nu!}\nonumber\\
   \nonumber\\
   &=2\sum_{\nu=2}^{\infty}\frac{t^{\nu}\Gamma^{\nu}\Omega^{\nu}}{N^{\nu}\nu!}.
\end{align}
Making use of Lemma F.2 from the supplementary information of \cite{childs2018toward}, we can bound the sum in (\ref{eq5-15}) as
\begin{align}
   \label{eq5-17}
   \dnorm{\exp(\frac{t}{N}\sum_{k=1}^{M}\gamma_{k}\mathcal{L}_{k})-\left(\sum_{k=1}^{M}p_{k}\exp(\omega\mathcal{L}_{k}) \right)}&\leq 2 \frac{t^{2}\Gamma^{2}\Omega^{2}}{2!N^{2}}\exp(\frac{t\Gamma\Omega}{N}).
\end{align}
Since $N$ is chosen such that $t\Gamma\Omega/N \leq 1$, we have $\exp(t\Gamma\Omega/N) \leq \exp(1) = e$, which gives the bound
\begin{align}
   \label{eq5-18}
   \dnorm{\exp(\frac{t}{N}\sum_{k=1}^{M}\gamma_{k}\mathcal{L}_{k})-\left(\sum_{k=1}^{M}p_{k}\exp(\omega\mathcal{L}_{k}) \right)}&\leq e\frac{t^{2}\Gamma^{2}\Omega^{2}}{N^{2}}.
\end{align}
Using the bound obtained in (\ref{eq5-18}) in the inequality (\ref{eq5-7}) yields
\begin{align}
   \label{eq5-19}
   \dnorm{T_{t}-\left(\mathcal{E}^{(QD)}_{\omega} \right)^{N}}&\leq N e\frac{t^{2}\Gamma^{2}\Omega^{2}}{N^{2}}=e\frac{t^{2}\Gamma^{2}\Omega^{2}}{N}.
\end{align}
Now choosing $\epsilon \geq 0$ such that
\begin{align}
\label{eq5-20}
   \epsilon \geq e\frac{t^{2}\Gamma^{2}\Omega^{2}}{N},
\end{align}
gives the desired bound
\begin{align}
   \label{eq5-21}
   \dnorm{T_{t}-\left(\mathcal{E}^{(QD)}_{\omega} \right)^{N}}&\leq \epsilon.
\end{align}
Also, from (\ref{eq5-20}) we have
\begin{align}
   N \geq e\frac{t^{2}\Gamma^{2}\Omega^{2}}{\epsilon}.
\end{align}
\end{proof}

\section{Implementation On A Quantum Computer}
\label{section6}
In the previous sections we have derived bounds for the error $\varepsilon$ and the number of steps $N$ for each of the randomised formulas as well as the QDRIFT channel. In this section we show how they can be implemented on a quantum computer using Classical Sampling (CS) to construct a gate set by sampling from a discrete distribution.

To implement the first order randomised product formula $S_{1}^{(ran)}(\tau)^{N}$, we need $N$ applications of $S_{1}^{(ran)}(\tau)$. However, the definition of $S_{1}^{(ran)}(\tau)$ in (\ref{eq3-2}) tells us that when we apply it to a state $\rho(0)$ it will apply the channel $S_{1}^{\rightarrow}(\tau)$ with probability $1/2$ and $S_{1}^{\leftarrow}(\tau)$ with probability $1/2$. This gives us a way to construct a gate set for $S_{1}^{(ran)}(\tau)$. If we define some random variable $j_{l} \in \{0,1\}$ for $l=1,...N$ where $p(j_{l}=0)=1/2=p(j_{l}=1)$ then, by assigning $S_{1}^{\rightarrow}(\tau)\equiv S_{1}^{(0)}(\tau)$ and $S_{1}^{\leftarrow}(\tau)\equiv S_{1}^{(1)}(\tau)$, we can construct a set of operations, denoted by $G_{1}^{(ran)}$, by iteratively sampling each $j_{l}$ and appending $S_{1}^{(j_{l})}(\tau)$ to $G_{1}^{(ran)}$. The gate set can then be applied to an initial state $\rho(0)$ to output the state $\tilde{\rho}(t)$ which approximates $\rho(t)$ to \rev{a} precision $\epsilon=e(Mt\Lambda)^{3}/3N^{2}$. The pseudo code for the algorithm that can construct this gate set is shown in Algorithm \ref{alg1}.

\begin{algorithm}[h!]
\caption{\rev{Classical sampling construction of the gate set $G_{1}^{(ran)}$ for the first order randomised Trotter-Suzuki product formula.}}
\label{alg1}
\textbf{Input}: A list of terms from the generator $\mathcal{L}=\sum_{k=1}^{M}\Lk{k}$ i.e. $\{\Lk{1},...,\Lk{M}\}$. A classical oracle function called SAMPLE() that returns a value $j_{l} \in \{0,1\}$ from the distribution $p(j_{l}=0)=1/2$ and $p(j_{l}=1)=1/2$. A target precision $\epsilon$ and a simulation time $t\geq 0$.
\\
\textbf{Output}: An ordered list $G_{1}^{(ran)}$ comprising of the product formulas $S_{1}^{\rightarrow} \equiv S_{1}^{(0)}$ and $S_{1}^{\leftarrow}\equiv S_{1}^{(1)}$.
\begin{algorithmic}[1]

\Require Compute $\Lambda := \max_{k}\dnorm{\Lk{k}}$ using the semidefinite program \cite{watrous2009semidefinite} to compute the diamond norm $\dnorm{\cdot}$.
\vspace{5mm}
\State $N \gets \left\lceil e^{1/2}(Mt\Lambda)^{3/2}/\sqrt{3\epsilon}\ \right\rceil$
\State $\tau \gets t/N$
\State $l \gets 0$
\State $G_{1}^{(ran)} =\{\}$ 
\While{$l < N$}
\State $j_{l} \gets$ SAMPLE()
\State Append $S_{1}^{(j_{l})}(\tau)$ to $G_{1}^{(ran)}$
\State $l \gets l+1$
\EndWhile
\State \textbf{return} $G_{1}^{(ran)}$
\end{algorithmic}
\end{algorithm}

\begin{figure*}[h]
\centering
\includegraphics[scale=0.5]{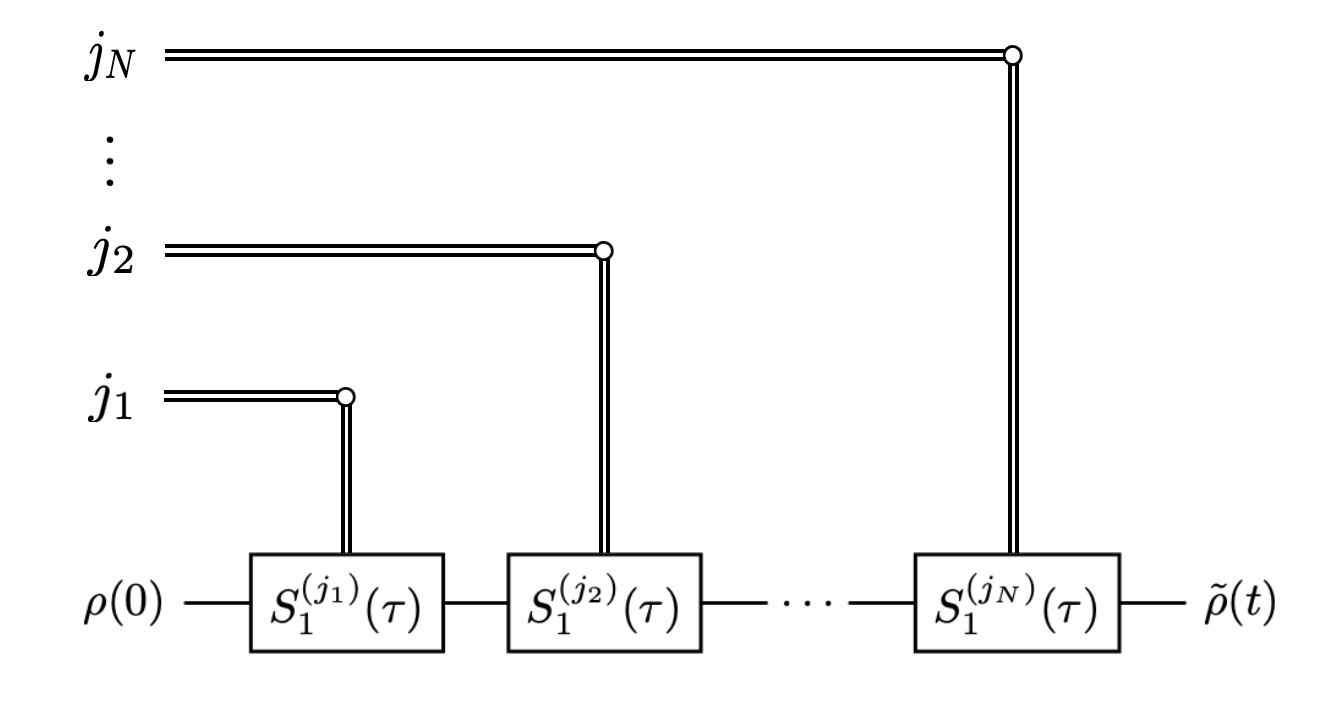}    
\caption{Quantum circuit showing the implementation of $S_{1}^{(ran)}(\tau)^{N}$ using CS. The circuit shows how the gate set $G_{1}^{(ran)}$ is constructed and applied to an initial state $\rho(0)$ with the output state $\tilde{\rho}(t)$. The double wires are used to show that the quantum channels depend on information from a classical computer. }
\label{Fig3}
\end{figure*}

\noindent The algorithm for implementing $S_{1}^{(ran)}(\tau)^{N}$ is shown in the circuit diagram in Figure \ref{Fig3}. In Figure \ref{Fig3} we see that each channel $S_{1}^{(j_{l})}$ depends on the value of each $j_{l}$ obtained from classical sampling.

\noindent We want to obtain the gate complexity for the circuit implemented by $G_{1}^{(ran)}$. To do this, we count the number of simple channels, which in this case is the number of exponentials,  $\exp(\tau\Lk{k})$, in the set of operations. Since there are $\lceil e^{1/2}(Mt\Lambda)^{3/2}/\sqrt{3\epsilon}\rceil$ TS product formulas in $G_{1}^{(ran)}$,  each containing $M$ exponentials, the gate complexity scales as $O(M^{5/2}(t\Lambda)^{3/2}/\sqrt{3\epsilon})$.

\begin{figure}[h]
   \centering
   \includegraphics[scale=0.5]{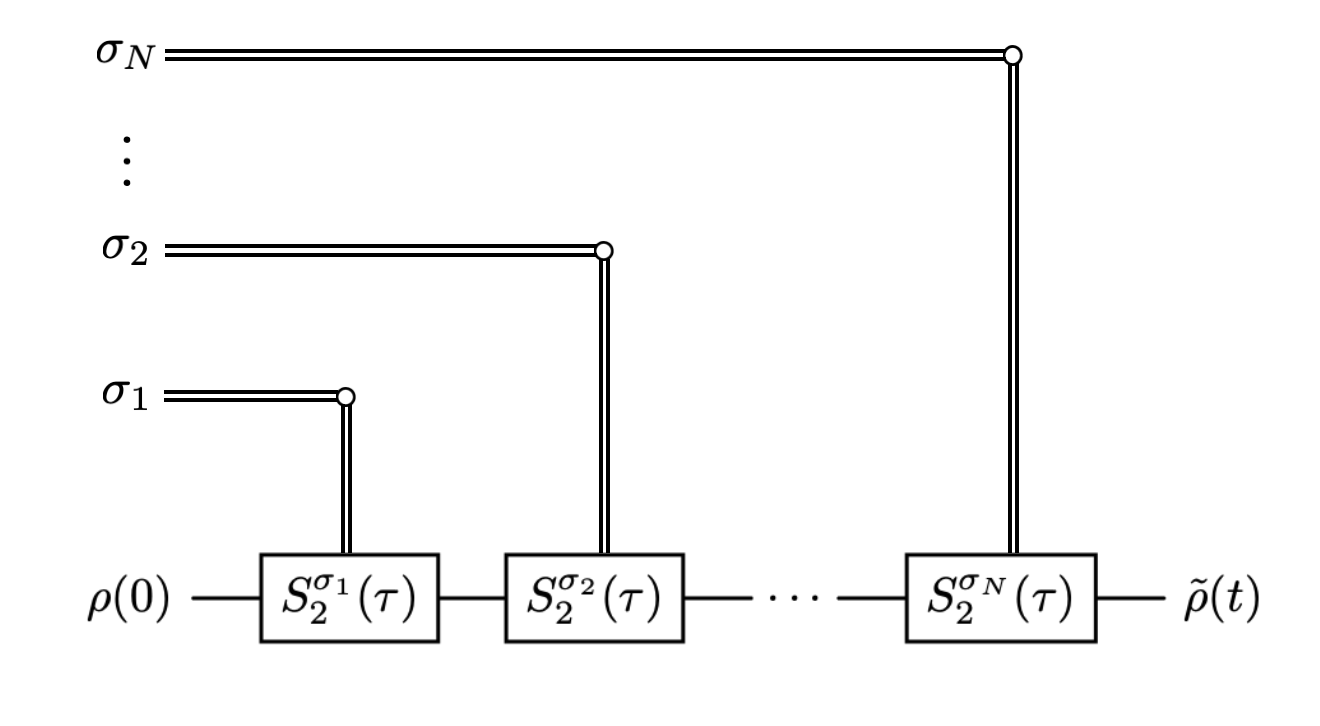}
   \caption{Quantum circuit showing the implementation of $S_{2}^{(ran)}(\tau)^{N}$ by CS. }
   \label{fig4}
\end{figure}

To implement the second order randomised formula $S_{2}^{(ran)}(\tau)^{N}$, we need $N$ applications of $S_{2}^{(ran)}(\tau)$ as in (\ref{eq4-6}). By definition $S_{2}^{(ran)}(\tau)$ applies $S_{2}^{\sigma}(\tau)$ for some permutation $\sigma \in \mathrm{Sym}(M)$ with a probability $1/M!$. But since the sum in (\ref{eq4-6}) is over all possible permutations in $\mathrm{Sym}(M)$ it is equally likely that we apply a $S_{2}^{\sigma}(\tau)$ for any permutation $\sigma$. So to construct a gate set that will implement $S_{2}^{(ran)}(\tau)^{N}$ we define $\sigma_{l}$, for $l=1,2,...,N$, which will be some permutation from $\mathrm{Sym}(M)$. Then we define an oracle function called SAMPLE\_PERMUTATION() which will sample from $\mathrm{Sym}(M)$, with every permutation in $\mathrm{Sym}(M)$ having a probability $1/M!$ of being sampled, and returning one of the permutations. We denote the gate set that we apply to initial state $\rho(0)$ as $G_2^{(ran)}$. Initially this set will be empty. Then, we iteratively obtain $\sigma_{l}$ using the function SAMPLE\_PERMUTATION() i.e. $\sigma_{l}=$SAMPLE\_PERMUTATION(), and append $S_{2}^{\sigma_{l}}$ to $G_{2}^{(ran)}$ repeating this process for $l=1,2,...,N$. Then, we can apply $G_{2}^{(ran)}$ to some initial sate $\rho(0)$ and the output will be the state $\tilde{\rho}(t)$ which approximates $\rho(t)$ to a precision $\epsilon=e(\Lambda t)^{3}M^{2}/N^{2}$. Algorithm \ref{alg2}, shows the pseudo code for how one can construct $G_{2}^{(ran)}$. Figure \ref{fig4}, shows the quantum circuit for applying each gate from $G_{2}^{(ran)}$  where each $\sigma_{l}$ is some permutation sampled from $\mathrm{Sym}(M)$. To see how the gate complexity for a circuit constructed with $G_{2}^{(ran)}$ scales, we count the number of exponentials. We will have $\lceil e^{1/2}(\Lambda t)^{3/2}M/\sqrt{\epsilon}\rceil$ TS product formulas in $G_{2}^{(ran)}$ and for each of these there are $2M$ exponentials. The gate complexity will then scale as $O((\Lambda t)^{3/2}M^{2}/\sqrt{\epsilon})$.

\begin{algorithm}[h!]
\caption{\rev{Classical sampling construction of the gate set $G_{2}^{(ran)}$ for the second order randomised Trotter-Suzuki product formula.}}
\label{alg2}
\textbf{Input}: A list of terms from the generator $\mathcal{L}=\sum_{k=1}^{M}\Lk{k}$ i.e. $\{\Lk{1},...,\Lk{M}\}$. A classical oracle function called SAMPLE\_PERMUTATION() that returns a permutation from $\mathrm{Sym}(M)$, with each permutation having the probability $1/M!$ of being sampled. A target precision $\epsilon$ and a simulation time $t\geq 0$.
\\
\textbf{Output}: An ordered list $G_{2}^{(ran)}$ comprising of the product formulas $S_{2}^{\sigma_{l}}$ where $\sigma_{l}\in \mathrm{Sym}(M)$.

\begin{algorithmic}[1]

\Require Compute $\Lambda := \max_{k}\dnorm{\Lk{k}}$ using the semidefinite program \cite{watrous2009semidefinite} to compute the diamond norm $\dnorm{\cdot}$.
\vspace{5mm}
\State $N \gets \left\lceil e^{1/2}(\Lambda t)^{3/2}M/\sqrt{\epsilon} \ \right\rceil$
\State $\tau \gets t/N$
\State $l \gets 0$
\State $G_{2}^{(ran)} =\{\}$
\While{$l < N$}
\State $\sigma_{l} \gets$ SAMPLE\_PERMUTATION()
\State Append $S_{2}^{(\sigma_{l})}(\tau)$ to $G_{2}^{(ran)}$
\State $l \gets l+1$
\EndWhile
\State \textbf{return} $G_{2}^{(ran)}$
\end{algorithmic}
\end{algorithm}

\begin{figure}[h]
   \centering
   \includegraphics[scale=0.35]{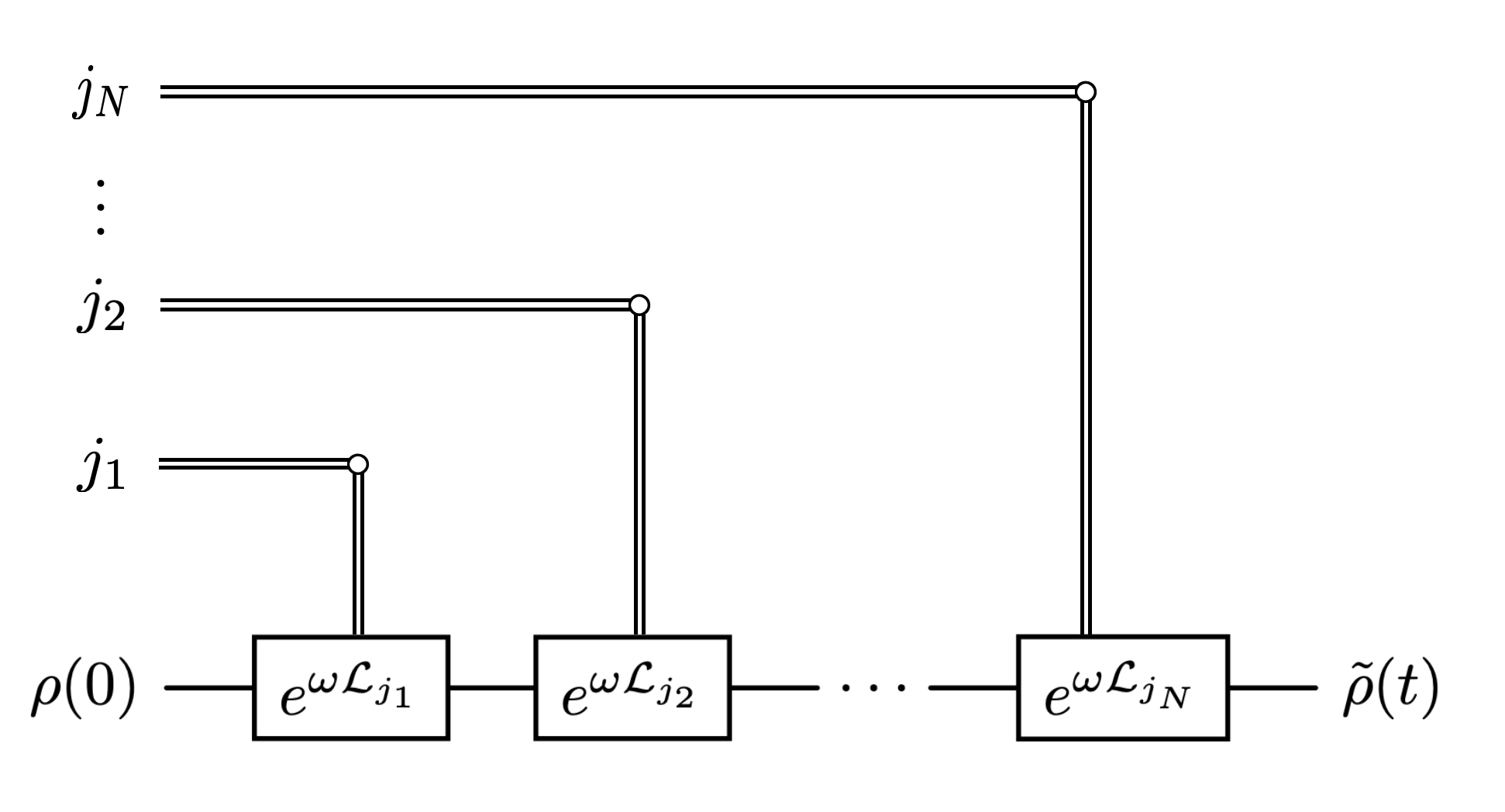}
   \caption{Quantum circuit depicting the implementation of the QDRIFT channel using CS. Here each $j_{k}$ is sampled from the discrete distribution $p_{k}$ as defined in equation (\ref{eq5-2}).}
   \label{fig5}
\end{figure}

\begin{algorithm}[h!]
\caption{\rev{Classical sampling construction of the gate set $G_{QD}$ for the QDRIFT channel.}}
\label{alg3}
\textbf{Input}: A list of terms from the generator $\mathcal{L}=\sum_{k=1}^{M}\gamma_{k}\mathcal{L}_{k}$ i.e. $\{\mathcal{L}_{1},...,\mathcal{L}_{M} \}$. A classical oracle function called SAMPLE() that returns a value $j_{l}$ from the probability distribution $p(j_{l}=k)=p_{k}=\gamma_{k}/ \Gamma$. A target precision $\epsilon$ and a simulation time $t\geq 0$.
\\
\textbf{Output}: An ordered list $G_{QD}$ comprising of the exponentials $\exp(\omega\mathcal{L}_{k})$.
\begin{algorithmic}[1]
\Require Compute $\Omega := \max_{k}\dnorm{\mathcal{L}_{k}}$ using the semidefinite program \cite{watrous2009semidefinite} to compute the diamond norm $\dnorm{\cdot}$.
\vspace{5mm}
\State $N \gets \left\lceil e(t\Gamma \Omega)^{2}/\epsilon\ \right\rceil$
\State $\omega \gets t\Gamma/N$
\State $l \gets 0$
\State $G_{QD} =\{\}$ 
\While{$l < N$}
\State $j_{l} \gets$ SAMPLE()
\State Append $\exp(\omega \mathcal{L}_{j_{l}})$ to $G_{QD}$
\State $l \gets l+1$
\EndWhile
\State \textbf{return} $G_{QD}$
\end{algorithmic}
\end{algorithm}

Consider the QDRFIT channel $\mathcal{E}_{\omega}^{(QD)}$ as in (\ref{eq5-3}). By definition, it will apply $\exp(\omega\mathcal{L}_{k})$ with some probability $p_{k}$ with $k=1,2,...,M$. This gives us an easy way to construct a quantum circuit that implements $(\mathcal{E}_{\omega}^{(QD)})^{N}$ using classical sampling. We start by defining a random variable $j_{l}\in \{1,...,M\}$, with $l=1,...,N$, and the probability distribution $p(j_{l}=k)=p_{k}$ with $k=1,...,M$. Then we denote the gate set by $G_{QD}$ and define the classical oracle function SAMPLE() which samples from the distribution $p(j_{l}=k)=p_{k}$ for $k=1,...,M$ and returns a value from the set $\{1,...,M\}$. To construct the circuit we iteratively use SAMPLE() to find a value $j_{l}$ and append $\exp(\omega \mathcal{L}_{j_{l}})$ for $l=1,...,N$ to the gate set $G_{QD}$. Once we have constructed the gate set, we apply it to the initial sate $\rho(0)$, to approximate $\rho(t)$ to a precision $\epsilon=e(t\Gamma \Omega)^{2}/N$. Algorithm \ref{alg3} outlines the pseudocode for constructing the gate set $G_{QD}$ to implement $(\mathcal{E}_{\omega}^{(QD)})^{N}$. Figure \ref{fig5}. shows the quantum circuit that implements the QDRIFT channel, we see here that each exponential depends on the outcome of sampling from the distribution $p_{k}$. Since the $G_{QD}$ contains only exponentials the gate complexity just scales here with the number of elements in $G_{QD}$ i.e. $O((t\Gamma\Omega)^{2}/\epsilon)$.

\section{Conclusion}
\label{section7}

We have shown that we can achieve faster simulation of Markovian OQS using randomisation. We have constructed randomised TS product formulas for simulating the dynamics of OQS. We have also proven directly that these formulas approximate the ideal evolution $T_{t}$. We  were able to prove all the results without using the Campbell and Hastings mixing Lemma \cite{campbell2017shorter,hastings2016turning}, which is vital to proving these results in the Hamiltonian simulation setting but not applicable to OQS simulation. With the first order randomised TS product formula, we achieve an improved precision, that is now quadratic in $t$, and we also have an improved gate complexity that scales the same as the second order deterministic TS product formula, where it scales with $M^{5/2}$.
For this formula we have provided a CS implementation whose gate complexity scales more efficiently in $M$ than the first order deterministic product formula. The second order randomised TS product formula has a much better scaling with respect to $M$ with the gate complexity depending quadratically on $M$. For the second order randomised TS product formula, we can only efficiently implement this using CS.
This work also proves that one can use the QDRIFT protocol \cite{campbell2019random} to simulate Markovian OQS. We see that for the quantum circuit implementation using CS, the gate complexity does not depend on $M$ making this method ideal for systems with many terms in the generator. However, the quadratic dependence on $t$ means that this method is only viable for short simulation times.
\\

\noindent It is important to contextualize our results within the broader landscape of Lindbladian simulation. As shown in our expanded comparison in Table \ref{Table1}, \rev{state-of-the-art deterministic algorithms such as the higher-order series expansion of Li and Wang \cite{li2022simulating}, and the repeated-interaction construction of Pocrnic \textit{et al.} \cite{pocrnic2023quantum} implemented via iterative qubitization, can achieve superior asymptotic scaling in simulation time $t$, approaching a near-linear dependence; the effective-Hamiltonian approach of Ding, Li and Lin \cite{ding2024simulating} likewise attains high-order accuracy through Hamiltonian simulation in an enlarged Hilbert space.} However, these advanced methods often rely on complex primitives like block-encodings or quantum signal processing, which can introduce significant constant-factor overheads in terms of gate counts and ancillary qubits. In the era of early fault-tolerant quantum computing, these overheads may be prohibitive. The randomized methods presented in this work, while having less favourable scaling in $t$, offer a compelling alternative due to their structural simplicity, lower implementation overhead, and direct composition from elementary channel operations. Therefore, they represent a practical and resource-efficient choice for near-term simulations, particularly for systems with a large number of generator terms $M$, where our randomized formulas provide a significant polynomial improvement over their direct deterministic counterparts.

\noindent A key technical contribution of this work is the direct derivation of error bounds for these randomised methods, which we achieved without relying on the Campbell and Hastings mixing lemma \cite{campbell2017shorter,hastings2016turning}. This direct approach was necessary because the mixing lemma is specific to the geometry of unitary operators and does not generalize to the CPTP maps that describe open system dynamics. While our Taylor expansion and bounding technique is general enough to be reapplied to the special case of Hamiltonian simulation, it would likely yield a more complex proof and potentially looser bounds than those achieved with the elegant mixing lemma. Given that tight bounds for randomised Hamiltonian simulation are already well-established, reapplying our more general method would not provide a clear benefit in that context. However, an interesting open problem that can be addressed is the generalisation of the mixing lemma for CPTP maps.

\noindent An open problem that can be addressed in future is to find optimal convex mixtures of second order TS product formulas that can improve precision and or acheive better scaling of the gate complexity in terms of $M$. Another open problem will be to see if one can apply these results to the simulation of quantum channels that describe non-Markovian dynamics of an OQS \cite{tamascelli2018nonperturbative,sweke2016digital, walters2024path}. Furthermore, a natural extension would be to develop randomised formulas of an order higher than two, analogous to the methods for Hamiltonian simulation \cite{childs2019faster}. The primary bottleneck for such an extension is the recursive construction of higher-order Suzuki formulas, which requires propagators with negative time steps. While evolving a Hamiltonian for a negative time corresponds to applying the adjoint unitary, the backward-in-time evolution generated by a Lindbladian, $e^{-t\mathcal{L}}$, is not in general a completely positive and trace-preserving (CPTP) map. This loss of physicality prevents a direct generalisation of higher-order randomised formulas to the open systems setting and overcoming it remains a significant challenge.

\section*{Author Contributions}

I. J. David, I. Sinayskiy and F. Petruccione jointly conceived the study and developed the
research idea. I. J. David derived the analytical results, constructed the proofs, and wrote
the manuscript. All authors discussed the results, and read and approved the final version of
the manuscript.

Artificial intelligence tools were used in the preparation of this manuscript, restricted to
copy-editing of the prose and to formatting of the \LaTeX{} source. They were not used to
generate, derive or verify any of the analytical results, proofs or bounds reported here, nor
to produce any of the figures. The authors have reviewed all such assisted output and take
full responsibility for the content of this work.

\begin{acknowledgements}

We would like to thank Ms. S. M. Pillay for her helpful discussions and assistance in proofreading the manuscript.
This work is based upon research supported by the
National Research Foundation of the Republic of South
Africa. Support from the NICIS (National Integrated
Cyber Infrastructure System) e-research grant QICSA is
kindly acknowledged.
Francesco Petruccione the Chair of the Scientific Board and Co-Founder of
QUNOVA computing. The authors declare no other competing interests.

\end{acknowledgements}

\bibliographystyle{quantum}
\bibliography{references}

\appendix
\numberwithin{equation}{section}
\section{Proofs Of Theorems and Lemmas In Section 2}
\label{AppendixA}
\begin{proof}{\textbf{(Lemma 1)}}
   We prove this by induction. For $N=0$ and $N=1$, the equality in equation (\ref{eq12}) holds and to show this would be trivial. So for the base case in our inductive proof, we choose $N=2$:
\begin{align}
       \label{eqA1}
       \dnorm{T^{2}-V^{2}}&=\dnorm{T^{2}-TV+TV-V^{2}}\nonumber\\
       &=\dnorm{T(T-V)+(T-V)V}\nonumber\\
       &\leq \dnorm{T(T-V)}+\dnorm{(T-V)V}\nonumber\\
       &\leq \dnorm{T}\dnorm{T-V}+\dnorm{T-V}\dnorm{V}
\end{align}
For any quantum channel $T$, by definition $\dnorm{T}=1$. This allows us to write
\begin{align}
       \label{eqA2}
       \dnorm{T^{2}-V^{2}}&\leq \dnorm{T-V}+\dnorm{T-V}=2\dnorm{T-V}.
\end{align}
   \\
Hence, we have verified that the inequality in equation (\ref{eq12}) holds for $N=2$. We now assume that it holds for $N=m$ and show that it is true for $N=m+1$:
\begin{align}
       \label{eqA3}
       \dnorm{T^{m+1}-V^{m+1}}&=\dnorm{T^{m+1}-TV^{m}+TV^{m}-V^{m+1}}\nonumber\\
       &=\dnorm{T(T^{m}-V^{m})+(T-V)V^{m}}\nonumber\\
       &\leq \dnorm{T}\dnorm{T^{m}-V^{m}}+\dnorm{T-V}\dnorm{V^{m}}\nonumber\\
       &\leq m\dnorm{T-V}+\dnorm{T-V}\dnorm{V}^{m}\\
       &\leq (m+1)\dnorm{T-V}.
\end{align}

   \noindent Therefore by induction the inequality in (\ref{eq12}) holds true for all integers $N\geq0$.
\end{proof}

\begin{proof}{(\textbf{Theorem 1.})}
To begin the proof, we define the parameter $\tau=t/N$ to be a small time step, and we recognise that
\begin{align}
\label{eqB1}
   \dnorm{T_{t}-S^{(det)}_{1}(t/N)^{N}}&=\dnorm{T_{t/N}^{N}-S^{(det)}_{1}(t/N)^{N}},\\
   &=\dnorm{T_{\tau}^{N}-S^{(det)}_{1}(\tau)^{N}},\\
   \label{eqB1(a)}
   &\leq N \dnorm{T_{\tau}-S^{(det)}_{1}(\tau)},
\end{align}

\noindent where the last inequality was obtained using Lemma \ref{lemma1}. The inequality above tells us that we need to find a bound on the distance between $T_{\tau}$ and $S_{1}^{(det)}(\tau)$. This is done by Taylor expanding both $T_{\tau}$ and $S_{1}^{(det)}(\tau)$ and looking at the remainder terms of their difference:
\begin{align}
   \label{eqB2}
   T_{\tau}-S_{1}^{(det)}(\tau)=\sum_{l=2}^{\infty}R_{l}(\tau)-W_{l}(\tau),\\
\end{align}

\noindent where $R_{l}(\tau)$ is the $l$-th order remainder term of the Taylor expansion of $T_{\tau}$ and $W_{l}(\tau)$ is the $l$-th order remainder term of the Taylor expansion of $S_{1}^{(det)}(\tau)$. Since $S_{1}^{(det)}(\tau)$ is first order it cancels with the first order terms in the Taylor expansion of $T_{\tau}$. Hence the remainder terms start from second order. Using equation (\ref{eqB2}), we find
\begin{align}
   \label{eqB3}
   \dnorm{T_{\tau}-S_{1}^{(det)}(\tau)}&=\dnorm{\sum_{l=2}^{\infty}R_{l}(\tau)-W_{l}(\tau)},\\
   &\leq \sum_{l=2}^{\infty}\dnorm{R_{l}(\tau)}+\dnorm{W_{l}(\tau)}.
\end{align}

\noindent To bound the term $R_{l}(\tau)$, we observe that it has the following expression:
\begin{align}
\label{eqB4}
   R_{l}(\tau)=\frac{\tau^{l}\mathcal{L}^{l}}{l!},
\end{align}
which allows us to write
\begin{align}
   \label{eqB5}
   \dnorm{R_{l}(\tau)}&=\dnorm{\frac{\tau^{l}\mathcal{L}^{l}}{l!}},\\
   &\leq \frac{\tau^{l}}{l!}\dnorm{\mathcal{L}}^{l},
\end{align}

\noindent where we have used the submultiplicativity of the diamond norm in the last line. To complete this bound we must bound the generator $\mathcal{L}$
\begin{align}
\label{eqB6}
   \dnorm{\mathcal{L}}=\dnorm{\sum_{k=1}^{M}\hat{\mathcal{L}}_{k}}\leq \sum_{k=1}^{M}\dnorm{\hat{\mathcal{L}}_{k}}\leq M\Lambda.
\end{align}
This allows us to complete the bound on $R_{l}(\tau)$ as
\begin{align}
   \label{eqB7}
   \dnorm{R_{l}(\tau)}\leq \frac{\tau^{l}M^{l}\Lambda^{l}}{l!}.
\end{align}

\noindent To bound the term $W_{l}(\tau)$ we need to find the general expression for the Taylor expansion of $S_{1}^{(det)})(\tau)$. Taylor expanding each exponential in equation (\ref{eq13}) leads to
\begin{align}
   \label{eqB8}
   S_{1}^{(det)}(\tau)=\sum_{j_{1},...,j_{M}=0}^{\infty} \frac{\tau^{j_{1}+...+j_{M}}}{j_{1}!...j_{M}!}\Lk{1}^{j_{1}}\Lk{2}^{j_{2}}...\Lk{M}^{j_{M}}.
\end{align}
The $M$ infinite sums can be written as one infinite sum and $M$ finite sums as
\begin{align}
\label{eqB9}
   \sum_{l=0}^{\infty}\sum_{\substack{j_{1},...,j_{M}=0;\\\sum_{\mu}j_{\mu}=l}}^{l} \frac{\tau^{j_{1}+...+j_{M}}}{j_{1}!...j_{M}!}\Lk{1}^{j_{1}}\Lk{2}^{j_{2}}...\Lk{M}^{j_{M}}.
\end{align}
The remainder term $W_{l}(\tau)$ is given by

\begin{align}
   \label{eqB10}
   W_{l}(\tau)=\sum_{\substack{j_{1},...,j_{M}=0;\\ \sum_{\mu}j_{\mu}=l}}^{l} \frac{\tau^{j_{1}+...+j_{M}}}{j_{1}!...j_{M}!}\Lk{1}^{j_{1}}\Lk{2}^{j_{2}}...\Lk{M}^{j_{M}}.
\end{align}
$W_{l}(\tau)$ can then be bounded as
\begin{align}
   \label{eqB11}
   \dnorm{W_{l}(\tau)}&\leq \sum_{\substack{j_{1},...,j_{M}=0;\\ \sum_{\mu}j_{\mu}=l}}^{l} \frac{\tau^{j_{1}+...+j_{M}}}{j_{1}!...j_{M}!}\dnorm{\Lk{1}}^{j_{1}}\dnorm{\Lk{2}}^{j_{2}}...\dnorm{\Lk{M}}^{j_{M}},\\
   \label{eqB11(a)}
   &\leq \sum_{\substack{j_{1},...,j_{M}=0;\\ \sum_{\mu}j_{\mu}=l}}^{l}\frac{\tau^{j_{1}+...+j_{M}}}{j_{1}!...j_{M}!}\Lambda^{j_{1}+...+j_{M}},
\end{align}
where the last inequality is obtained by the fact that $\dnorm{\Lk{k}}\leq \Lambda$ for all $k=1,...,M$. To complete the bound we must compute the restricted sum in equation (\ref{eqB11(a)}). We make use of Lemma \ref{lemmaRestrictedSum} found in Appendix \ref{AppendixB} to do this which leads to,
\begin{align}
   \label{eqB12}
   \dnorm{W_{l}(\tau)}\leq \frac{\tau^{l}M^{l}\Lambda^{l}}{l!}.
\end{align}
Using equations (\ref{eqB7}) and (\ref{eqB12}), we can bound the distance
\begin{align}
   \label{eqB13}
   \dnorm{T_{\tau}-S^{(det)}_{1}(\tau)}&\leq \sum_{l=2}^{\infty}\dnorm{R_{l}(\tau)}+\dnorm{W_{l}(\tau)},\\
   &\leq \sum_{l=2}^{\infty} \frac{\tau^{l}M^{l}\Lambda^{l}}{l!} + \frac{\tau^{l}M^{l}\Lambda^{l}}{l!},\\
   \label{eqB13(a)}
   &=2 \sum_{l=2}^{\infty} \frac{\tau^{l}M^{l}\Lambda^{l}}{l!}.
\end{align}
Making use of Lemma F.2 from the supplementary information of \cite{childs2018toward}, which states that for some $y \geq 0 \in \mathbb{R}$ and $k \in \mathbb{N}$,
\begin{align}
   \label{eqB14}
   \sum_{n=k}\frac{y^n}{n!}\leq \frac{y^{k}}{k!}\exp(y),
\end{align}
we can bound the sum in equation (\ref{eqB13(a)}) as
\begin{align}
   \label{eqB15}
   \dnorm{T_{\tau}-S^{(det)}_{1}(\tau)}&\leq \frac{2\tau^{2}\Lambda^{2}M^{2}}{2}\exp(\tau\Lambda M),\\
   & =\frac{t^{2}\Lambda^{2}M^{2}}{N^{2}}\exp(\frac{t\Lambda M}{N}),
\end{align}
where we have replaced $\tau$ by $t/N$. Since $N$ is chosen such that $Mt\Lambda/N \leq 1$, we have $\exp(t\Lambda M/N) \leq \exp(1) = e$, which gives
\begin{align}
   \label{eqB16}
   \dnorm{T_{\tau}-S^{(det)}_{1}(\tau)}\leq e\frac{t^{2}\Lambda^{2}M^{2}}{N^{2}}.
\end{align}
Substituting equation (\ref{eqB16}) into equation (\ref{eqB1(a)}), we get the desired result
\begin{align}
   \label{eqB17}
   \dnorm{T_{t}-S^{(det)}_{1}(t/N)^{N}}\leq N  e\frac{t^{2}\Lambda^{2}M^{2}}{N^{2}} =  e\frac{t^{2}\Lambda^{2}M^{2}}{N}.
\end{align}
\end{proof}

\begin{proof}{(\textbf{Theorem 2.})}
Theorem 2 is proved in a similar manner to Theorem 1. Start by defining $\tau=t/N$ and applying Lemma \ref{lemma1}. to equation (\ref{eq17}),
\begin{align}
   \label{eqC1}
   \dnorm{T_{t}-S_{2}^{(det)}(\tau)^{N}}\leq N\dnorm{T_{\tau}-S_{2}^{(det)}(\tau)}.
\end{align}

\noindent The distance $\dnorm{T_{\tau}-S_{2}^{(det)}(\tau)}$ is bounded by considering the remainder terms in the Taylor expansions of the difference $T_{\tau}-S_{2}^{(det)}(\tau)$. Similarly to the proof of theorem 1 we let $R_{l}(\tau)$ and $W_{l}(\tau)$ be the $l$-th order remainder terms of the Taylor expansions of $T_{\tau}$ and $S_{2}^{(det)}(\tau)$ respectively. This leads to
\begin{align}
   \label{eqC2}
   T_{\tau}-S_{2}^{(det)}(\tau)=\sum_{l=3}^{\infty}R_{l}(\tau)-W_{l}(\tau).
\end{align}
Here, the index $l$ starts from 3 since $S_{2}^{(det)}(\tau)$ is second order accurate. This is what it means for $S_{2}^{(det)}(\tau)$ to be a second order product formula. Using the subadditivity of the diamond norm, we obtain
\begin{align}
   \label{eqC3}
   \dnorm{T_{\tau}-S_{2}^{(det)}(\tau)}\leq \sum_{l=3}^{\infty}\dnorm{R_{l}(\tau)}+\dnorm{W_{l}(\tau)}.
\end{align}

\noindent The term $R_{l}(\tau)$ has the same form in equation (\ref{eqB4}) and so has the same bound in equation (\ref{eqB7}). To find the bound on $\dnorm{W_{l}(\tau)}$ we need to find a general expression for the Taylor expansion of $S_{2}^{(det)}(\tau)$. This is found to be
\begin{align}
   \label{eqC4}
   S_{2}^{(det)}(\tau)=\sum_{\substack{j_{1},...,j_{M}=0\\k_{1},...,k_{M}=0}}^{\infty}\frac{(\tau/2)^{j_{1}+...+j_{M}+k_{1}+...+k_{M}}}{j_{1}!...j_{M}!k_{1}!...k_{M}!}\prod_{q=1}^{M}\Lk{q}^{j_{q}}\prod_{p=M}^{1}\Lk{p}^{k_{p}}.
   \end{align}
This can be rewritten to only contain one infinite sum as
\begin{align}
   \label{eqC5}
   S_{2}^{(det)}(\tau)=\sum_{l=0}^{\infty}\sum_{\substack{j_{1},...,j_{M}=0\\k_{1},...,k_{M}=0\\ \sum_{\mu}j_{\mu}+\sum_{\nu}k_{\nu}=l}}^{l}\frac{(\tau/2)^{ \sum_{\mu}j_{\mu}+\sum_{\nu}k_{\nu}}}{j_{1}!...j_{M}!k_{1}!...k_{M}!}\prod_{q=1}^{M}\Lk{q}^{j_{q}}\prod_{p=M}^{1}\Lk{p}^{k_{p}}.
\end{align}
This allows us to write the remainder term as
\begin{align}
   \label{eqC6}
   W_{l}(\tau)=\sum_{\substack{j_{1},...,j_{M}=0\\k_{1},...,k_{M}=0\\ \sum_{\mu}j_{\mu}+\sum_{\nu}k_{\nu}=l}}^{l}\frac{(\tau/2)^{ \sum_{\mu}j_{\mu}+\sum_{\nu}k_{\nu}}}{j_{1}!...j_{M}!k_{1}!...k_{M}!}\prod_{q=1}^{M}\Lk{q}^{j_{q}}\prod_{p=M}^{1}\Lk{p}^{k_{p}}.
\end{align}
Then subadditivity of the diamond norm leads to the bound
\begin{align}
   \label{eqC7}
   \dnorm{W_{l}(\tau)}\leq \sum_{\substack{j_{1},...,j_{M}=0\\k_{1},...,k_{M}=0\\ \sum_{\mu}j_{\mu}+\sum_{\nu}k_{\nu}=l}}^{l}\frac{(\tau/2)^{ \sum_{\mu}j_{\mu}+\sum_{\nu}k_{\nu}}}{j_{1}!...j_{M}!k_{1}!...k_{M}!}\dnorm{\prod_{q=1}^{M}\Lk{q}^{j_{q}}\prod_{p=M}^{1}\Lk{p}^{k_{p}}}.
\end{align}

\noindent We now need to bound the diamond norm of the product of the summands $\Lk{k}$ in equation (\ref{eqC7}). This is done using submultiplicativity of the diamond norm as
\begin{align}
   \label{eqC8}
   \dnorm{\prod_{q=1}^{M}\Lk{q}^{j_{q}}\prod_{p=M}^{1}\Lk{p}^{k_{p}}}&\leq \prod_{q=1}^{M}\dnorm{\Lk{q}^{j_{q}}}\prod_{p=M}^{1}\dnorm{\Lk{p}^{k_{p}}},\\
   &\leq \prod_{q=1}^{M}\dnorm{\Lk{q}}^{j_{q}}\prod_{p=M}^{1}\dnorm{\Lk{p}}^{k_{p}},\\
   \label{eqC8(a)}
   &\leq \Lambda^{j_{1}+...j_{M}+k_{1}+...+k_{M}},
\end{align}

\noindent where the last inequality is obtained using the fact that $\dnorm{\Lk{k}}\leq \Lambda$ for all $k$. Using the inequality (\ref{eqC8(a)}) with equation (\ref{eqC7}) yields
\begin{align}
   \label{eqC9}
   \dnorm{W_{l}(\tau)}\leq \sum_{\substack{j_{1},...,j_{M}=0\\k_{1},...,k_{M}=0\\ \sum_{\mu}j_{\mu}+\sum_{\nu}k_{\nu}=l}}^{l}\frac{(\tau/2)^{ \sum_{\mu}j_{\mu}+\sum_{\nu}k_{\nu}}}{j_{1}!...j_{M}!k_{1}!...k_{M}!} \Lambda^{\sum_{\mu}j_{\mu}+\sum_{\nu}k_{\nu}}.
\end{align}
We can compute this restricted sum by using Lemma B.1. with the following replacements: $x\rightarrow \tau\Lambda/2$, $p\rightarrow l$ and $M\rightarrow 2M$. This leads to
\begin{align}
\label{eqC10}
   \dnorm{W_{l}(\tau)}\leq \frac{2^{l}M^{l} \ \tau^{l} \ \Lambda^{l}}{2^{l} \ l!}=\frac{M^{l} \ \tau^{l} \ \Lambda^{l}}{l!}.
\end{align}
Now using equations (\ref{eqB7}) and (\ref{eqC10}), we can bound the norm as
\begin{align}
   \label{eqC11}
   \dnorm{T_{\tau}-S_{2}^{(det)}(\tau)}&\leq \sum_{l=3}^{\infty} \dnorm{R_{l}(\tau)}+\dnorm{W_{l}(\tau)}\\
   &\leq \sum_{l=3}^{\infty}\frac{M^{l}\tau^{l}\Lambda^{l}}{l!}+\frac{M^{l} \ \tau^{l} \ \Lambda^{l}}{l!}\\
   &=2\sum_{l=3}^{\infty}\frac{M^{l} \ \tau^{l} \ \Lambda^{l}}{l!}.
\end{align}
Using equation (\ref{eqB14}), we can write the bound above as
\begin{align}
   \label{eqC12}
   \dnorm{T_{\tau}-S_{2}^{(det)}(\tau)}&\leq 2 \frac{M^{3}\tau^{3}\Lambda^{3}}{3!}\exp(M\tau\Lambda)\\
   &= \frac{M^{3}t^{3}\Lambda^{3}}{3N^{3}}\exp(\frac{Mt\Lambda}{N}).
\end{align}
Since $N$ is chosen such that $Mt\Lambda/N \leq 1$, we have $\exp(Mt\Lambda/N) \leq \exp(1) = e$, which yields
\begin{align}
   \label{eqC13}
   \dnorm{T_{\tau}-S_{2}^{(det)}(\tau)}&\leq  e\frac{M^{3}t^{3}\Lambda^{3}}{3N^{3}}.
\end{align}
Now using equations (\ref{eqC13}) and (\ref{eqC1}) we can bound the norm as
\begin{align}
   \label{eqC14}
   \dnorm{T_{t}-S_{2}^{(det)}\left(\frac{t}{N}\right)^{N}}\leq N \ e\frac{M^{3}t^{3}\Lambda^{3}}{3N^{3}}= e\frac{M^{3}t^{3}\Lambda^{3}}{3N^{2}}.
\end{align}
Then letting $\epsilon \geq 0$ and
\begin{align}
   \label{eqC15}
   \epsilon \geq e\frac{M^{3}t^{3}\Lambda^{3}}{3N^{2}}
\end{align}
then we have that
\begin{align}
   \label{eqC16}
   \dnorm{T_{t}-S_{2}^{(det)}\left(\frac{t}{N}\right)^{N}}\leq \epsilon.
\end{align}
Given the bound on the precision $\epsilon$, we can find a bound on $N$ as
\begin{align}
   \label{eqC17}
   N\geq e^{1/2}\frac{M^{3/2}t^{3/2}\Lambda^{3/2}}{(3\epsilon)^{1/2}},
\end{align}
completing the proof.
\end{proof}

\section{Results For Computing Restricted Sums}
\label{AppendixB}
The following result allows us to compute restricted sums.

\begin{lemma}
\label{lemmaRestrictedSum}
   Given some $p,M\in \mathbb{N}$ and some $x\in \mathbb{R}$,
   \begin{align}
       \sum_{\substack{j_{1},...,j_{M}=0;\\ \sum_{\mu}j_{\mu}=p}}^{p}\frac{x^{j_{1}+...+j_{M}}}{j_{1}!...j_{M}!}=\frac{M^{p}x^{p}}{p!}.
   \end{align}
\end{lemma}

\begin{proof}
   Consider the exponential function $\exp(x)$ for some real number $x$,
\begin{align}
   e^{Mx}=\underbrace{e^{x}e^{x}...e^{x}}_{M\text{ - times}}.
\end{align}
By Taylor expanding each exponential on the left hand side of equation (\ref{eqB12}), we get
\begin{align}
   \underbrace{e^{x}e^{x}...e^{x}}_{M\text{ - times}}&=\sum_{j_{1},...,j_{M}=0}^{\infty}\frac{x^{j_{1}+...+j_{M}}}{j_{1}!...j_{M}!},\\
   \label{eqressum1}
   &=\sum_{p=0}^{\infty}\sum_{\substack{j_{1},...,j_{M}=0;\\ \sum_{\mu}j_{\mu}=p}}^{p}\frac{x^{j_{1}+...+j_{M}}}{j_{1}!...j_{M}!}.
\end{align}
But, $\exp(x)^{M}=\exp(Mx)$ has the following Taylor expansion
\begin{align}
\label{eqressum2}
   \exp(Mx)=\sum_{p=0}^{\infty}\frac{M^{p}x^{p}}{p!}.
\end{align}
Equating equations (\ref{eqressum1}) and (\ref{eqressum2}) and comparing terms leads us to the desired result
\begin{align}
   \sum_{\substack{j_{1},...,j_{M}=0;\\ \sum_{\mu}j_{\mu}=p}}^{p}\frac{x^{j_{1}+...+j_{M}}}{j_{1}!...j_{M}!}=\frac{M^{p}x^{p}}{p!}.
\end{align}
\end{proof}

\end{document}